%Paper: astro-ph/9310048
%From: stebbins@perseus.fnal.gov (Albert Stebbins)
%Date: Wed, 27 Oct 93 22:58:02 -0500

%
%	This TeX file uses epsf.tex which is available from the bulliten board
%       via "get epsf.tex".  If you are unable to include PostScript graphics 
%	with TeX you may want to commment out all use of \putfig and \putfigp
%       below.
%
%	Six uuencoded, compressed, tarred PostScript files are included at the
%       end of this file, below the line "cut here".
%       

\magnification=\magstep1
\baselineskip=16pt
\hoffset=-0.75truecm
\voffset=0.0truecm
\vsize=23.5truecm
\hsize=18.0truecm
\parskip=0.2cm
\parindent=1cm

\def \header#1{\goodbreak\bigskip\centerline{\bf #1}\medskip\nobreak}
\def \subheader#1{\goodbreak\medskip\par\noindent{\bf #1}\smallskip\nobreak}

\def \etal {{\it et al.} }

\def\s {\scriptscriptstyle}

\def\sqr#1#2{{\vcenter{\hrule height.#2pt
              \hbox{\vrule width.#2pt height#1pt \kern#1pt \vrule width.#2pt}
\hrule height.#2pt}}}
\def\square{\mathchoice\sqr59\sqr59\sqr46\sqr36}
\def\mathrelfun#1#2{\lower3.6pt\vbox{\baselineskip0pt\lineskip.9pt
  \ialign{$\mathsurround=0pt#1\hfil##\hfil$\crcr#2\crcr\sim\crcr}}}

\def\rmf {{\rm f}}

\def\rmp {{\rm p}}

\def\rmK {{\rm K}}

\def\bfn {{\bf n}}

\def\calD {{\cal D}}

\def\calF {{\cal F}}

\def\calN {{\cal N}}

\def\calP {{\cal P}}

\def\calS {{\cal S}}
\def\calT {{\cal T}}

\def\hatbfn  {{\hat\bfn}}

\def\hatn{{\hat n}}

\def\eV  {{\rm \hbox{e\kern-0.14em V}}}
\def\keV {{\rm \hbox{ke\kern-0.14em V}}}
\def\MeV {{\rm \hbox{Me\kern-0.14em V}}}
\def\GeV {{\rm \hbox{Ge\kern-0.14em V}}}

\input epsf

% Received from Jnet%SLACVM::SHARON  1-MAR-1988 13:16
% Subj:	TABLES TEX
% +--------------------------------------------------------------------+
% |                                                                    |
% |                           TABLES.TEX                               |
% |                                                                    |
% |                     Ray F. Cowan  15-Feb-85                        |
% |                                                                    |
% |                       Princeton University                         |
% |                                                                    |
% |                     Last Revision: 17-Apr-86                       |
% |                                                                    |
% |   Macros I find handy for making tables.  See TABLEDOC TEX for     |
% |   a longer description.  The token-counting macros are straight    |
% |   from the TeXbook's "Dirty Tricks" appendix.                      |
% |                                                                    |
% +--------------------------------------------------------------------+
%
%AJS3 renamed "\header" macro to "\headertab" to avoid interference with other
%     "\header" macros.
%
\newbox\hdbox%
\newcount\hdrows%
\newcount\multispancount%
\newcount\ncase%
\newcount\ncols% This is the number of primary text columns in the table.
\newcount\nrows%
\newcount\nspan%
\newcount\ntemp%
\newdimen\hdsize%
\newdimen\newhdsize%
\newdimen\parasize%
\newdimen\spreadwidth%
\newdimen\thicksize%
\newdimen\thinsize%
\newdimen\tablewidth%
\newif\ifcentertables%
\newif\ifendsize%
\newif\iffirstrow%
\newif\iftableinfo%
\newtoks\dbt%
\newtoks\hdtks%
\newtoks\savetks%
\newtoks\tableLETtokens%
\newtoks\tabletokens%
\newtoks\widthspec%
%
%  Book-keeping stuff--see how often these macros are called.
%
\immediate\write15{%
CP SMSG GJMSINK TEXTABLE --> TABLE MACROS V. 851121 JOB = \jobname%
}%
%
%  Turn on table diagnostics.
%
\tableinfotrue%
\catcode`\@=11%  Allows use of "@" in macro names, like PLAIN.TEX does.
%  Debugging aid.  Writes #1 on the
%                                    user's terminal and in the log file.
%
%  Define the \tstrut height, depth in terms of the x_height parameter.
%
\def\tstrut{\vrule height3.1ex depth1.2ex width0pt}%
\def\and{\char`\&}%  Allows us to get an `&' in the text.  This is the
%                    same as using the PLAIN TeX macro \&.
\def\tablerule{\noalign{\hrule height\thinsize depth0pt}}%
\thicksize=1.5pt%  Default thickness for fat rules.  The user should feel
%                  free to change this to his preference.
\thinsize=0.6pt%   Default thickness for thin rules.
\def\thickrule{\noalign{\hrule height\thicksize depth0pt}}%
\def\ctr#1{\hfil\ #1\hfil}%
%
%
%
%  Here are things for controlling the width of the finished table.
%
\tablewidth=-\maxdimen%
\spreadwidth=-\maxdimen%
\def\tabskipglue{0pt plus 1fil minus 1fil}%
%
%  Stuff for centering or not.
%
\centertablestrue%
%
%
%
%  \vctr vertically centers its argument in the row.
%
\parasize=4in%
\gdef\ARGS{########}%  Produces the correct number of #'s in the preamble
%                      by the time eveything is expanded and \halign sees
%                      it.
\gdef\headerARGS{####}%  Same as \ARGS, but used in \headertab macros.
\def\@mpersand{&}%  Allows us to get alignment tab characters later
%                   when we have made the character "&" an active macro.
{\catcode`\|=13%  Make |'s locally active.
\gdef\letbarzero{\let|0}%  Globally define a macro that allows us to
%                          keep active |'s from being expanded in edef's.
\gdef\letbartab{\def|{&&}}%
\gdef\letvbbar{\let\vb|}%
%  This \def will cause active |'s read by
%                            \ruledtable to be converted into double
%                            alignment tabs.
}%  End of locally active |'s.
{\catcode`\&=4%  Make these alignment tabs.
\def\ampskip{&\omit\hfil&}%  This local macro skips a vertical rule.
\catcode`\&=13%  Now make &'s into active macros.
\let&0%  This allows us to expand \ampskip in the next \xdef without
%        attempting to expand the & and getting an "undefined control
%        sequence" error.
\xdef\letampskip{\def&{\ampskip}}%
\gdef\letnovbamp{\let\novb&\let\tab&}
%  This will cause active &'s read by
%                                   \ruledtable to be converted into
%                                   double tabs and an \omit'ted \vrule.
}%  End of locally active &'s.
\def\begintable{%  Here we make |'s and &'s active characters so we can
%                  interpret them as macros.  Note that this action is
%                  true only until we encounter the matching \endgroup
%                  token later at the end of the \ruledtable macro.
   \begingroup%
   \catcode`\|=13\letbartab\letvbbar%
   \catcode`\&=13\letampskip\letnovbamp%
   \def\multispan##1{%  We must redefine \multispan to count the number
%                       of primary columns, not physical columns.
      \omit \mscount##1%
      \multiply\mscount\tw@\advance\mscount\m@ne%
      \loop\ifnum\mscount>\@ne \sp@n\repeat%
   }%  End of \multispan macro.
   \def\|{%
      &\omit\widevline&%
   }%
   \ruledtable%  Now we call \ruledtable to do the real work.
}%  End of \begintable macro.
\long\def\ruledtable#1\endtable{%
%
%  This macro reads in the user's data entries
%  and converts them into a ruled table.
%
%  Important note:  Many macros and parameters are re-defined here, and
%  these must be kept local to the table macros to avoid conflict with
%  their use outside of tables.  This is done by the \begingroup token
%  macro \begintable and the \endgroup token at the end of
%  this macro.
%
   \offinterlineskip%  Needed to make rules touch each other.
   \tabskip 0pt%  Needed for same reason as \offinterlineskip.
   \def\widevline{\vrule width\thicksize}%  Make outer \vrule's wider.
   \def\endrow{\@mpersand\omit\hfil\crnorm\@mpersand}%
   \def\crthick{\@mpersand\crnorm\thickrule\@mpersand}%
   \def\crthickneg##1{\@mpersand\crnorm\thickrule
          \noalign{{\skip0=##1\vskip-\skip0}}\@mpersand}%
   \def\crnorule{\@mpersand\crnorm\@mpersand}%
   \def\crnoruleneg##1{\@mpersand\crnorm
          \noalign{{\skip0=##1\vskip-\skip0}}\@mpersand}%
   \let\nr=\crnorule%  A shorter abbreviation.
   \def\endtable{\@mpersand\crnorm\thickrule}%
   \let\crnorm=\cr%  Allows us to use \cr for our own purposes.
%
%  Cause user-typed \cr's to follow a row with a \tablerule.
%
   \edef\cr{\@mpersand\crnorm\tablerule\@mpersand}%
   \def\crneg##1{\@mpersand\crnorm\tablerule
          \noalign{{\skip0=##1\vskip-\skip0}}\@mpersand}%
   \let\ctneg=\crthickneg
   \let\nrneg=\crnoruleneg
   \the\tableLETtokens%  Get the user's extra \let's, if any.
%
%  Put the data entries into a token register so we can scan through them
%  and see what the user is asking us to do.
%
   \tabletokens={&#1}%  We add an extra alignment tab to the beginning
%                       of the first row to allow for the first \vrule.
%
%  Now count how many rows are in the table and return the result in
%  count register \nrows; do the same for columns, and return that
%  in register \ncols.
%
   \countROWS\tabletokens\into\nrows%
   \countCOLS\tabletokens\into\ncols%
%
%  Now do a little arithmetic to convert the number of primary columns
%  into the number of physical columns that the alignment preamble must
%  prepare for;  similarly for rows.
%
   \advance\ncols by -1%
   \divide\ncols by 2%
   \advance\nrows by 1%
%
%  Tell the user how many rows and columns we found in his data, if he
%  wants to know.
%
   \iftableinfo %
      \immediate\write16{[Nrows=\the\nrows, Ncols=\the\ncols]}%
   \fi%
%
%  Now we actually go ahead and produce the table.
%
   \ifcentertables
      \ifhmode \par\fi%  Make sure we are in vertical mode.
      \line{%  The final table comes out as an \hbox of width the \hsize.
      \hss%  The final table will be centered left-to-right.
   \else %
      \hbox{%
   \fi
      \vbox{%
         \makePREAMBLE{\the\ncols}%  Generate the preamble.
         \edef\next{\preamble}%  This line and the next line force the
         \let\preamble=\next%    expansion of all \ARGS tokens into the
%                                appropriate number of #'s.
         \makeTABLE{\preamble}{\tabletokens}%  Go do the \halign here.
      }%  End of \vbox.
      \ifcentertables \hss}\else }\fi%  Finish the centering effect.
%                                       It is important that no spaces
%                                       follow the two `}' here.
%  }%  End of \line.
   \endgroup%  Return all local macros and parameters to their outside
%              values.
   \tablewidth=-\maxdimen%  Reset \tablewidth to normal.
   \spreadwidth=-\maxdimen% Same for \spreadwidth.
}%  End of macro \ruledtable.
\def\makeTABLE#1#2{%  Does an \halign for the \ruledtable macro.
   {%  Start of local parameter values.
   \let\ifmath0%     These macros would cause trouble if they were to be
   \let\headertab0%  expanded in the following \xdef; we \let them be
   \let\multispan0%  equal to a digit, because digits can't be expanded.
%
%  Set up the width specification here.
%
   \ncase=0%
   \ifdim\tablewidth>-\maxdimen \ncase=1\fi%
   \ifdim\spreadwidth>-\maxdimen \ncase=2\fi%
   \relax%  This \relax is absolutely necessary, without it the following
%           \ifcase will always take \ncase=0.
%
   \ifcase\ncase %
      \widthspec={}%
   \or %
      \widthspec=\expandafter{\expandafter t\expandafter o%
                 \the\tablewidth}%
   \else %
      \widthspec=\expandafter{\expandafter s\expandafter p\expandafter r%
                 \expandafter e\expandafter a\expandafter d%
                 \the\spreadwidth}%
   \fi %
%\out{Widthspec=[\the\widthspec]}%
%\out{Preamble=[\preamble]}%
   \xdef\next{%  We must force the preamble to be expanded BEFORE the
      \halign\the\widthspec{%
%        \halign is done;  this \edef\next{...}\next construction
%                does the trick.
      #1%  This is the preamble text.
      \noalign{\hrule height\thicksize depth0pt}%  Makes the top \hrule.
      \the#2\endtable%  This is the main body.
%
%     \noalign{\hrule height0.7pt depth0pt}%  Makes the last \hrule.
      }%  End of \halign.
   }%  End of \next.
   }%  End of local values.
   \next%  This \next must be outside of the local values, because now
%          we want those troublesome macros in the \let's above to have
%          their normal actions.
}%  End of macro \makeTABLE.
\def\makePREAMBLE#1{%  This macro generates the necessary preamble for a
%                      ruled table with #1 primary columns.
%                      (Primary columns means the number of columns NOT
%                       counting those used for vertical rules.)
   \ncols=#1%  Get the number of columns desired.
   \begingroup%  Start local parameter definitions.
   \let\ARGS=0%  This is the key to the whole thing; it prevents \ARGS
%                from being expanded in the following \edef's.
   \edef\xtp{\widevline\ARGS\tabskip\tabskipglue%
   &\ctr{\ARGS}\tstrut}%  A 1-column preamble.  Gets the sizing right.
   \advance\ncols by -1%  One column has been generated; decrement the
%                         counter.
   \loop%  Append as many further columns as needed to the preamble.
      \ifnum\ncols>0 %
      \advance\ncols by -1%
      \edef\xtp{\xtp&\vrule width\thinsize\ARGS&\ctr{\ARGS}}%
   \repeat
   \xdef\preamble{\xtp&\widevline\ARGS\tabskip0pt%
   \crnorm}%  Adds the last \vrule.
   \endgroup%  End of local parameters.
}%  End of macro \makePREAMBLE.
\def\countROWS#1\into#2{%  This counts the number of rows in #1 by
%                          looking for control sequences that end a row,
%                          e.g., \cr, \crthick, etc., and puts the result
%                          into count register #2.
   \let\countREGISTER=#2%
   \countREGISTER=0%
%  \out{In countROWS:  tokens are [\the#1]}%
   \expandafter\ROWcount\the#1\endcount%
}%
\def\ROWcount{%
   \afterassignment\subROWcount\let\next= %
}%
\def\subROWcount{%
%  \out{In subROWcount:  next is [\meaning\next]}%  Debugging aid.
   \ifx\next\endcount %
      \let\next=\relax%
   \else%
      \ncase=0%
      \ifx\next\cr %
         \global\advance\countREGISTER by 1%
         \ncase=0%
      \fi%
      \ifx\next\endrow %
         \global\advance\countREGISTER by 1%
         \ncase=0%
      \fi%
      \ifx\next\crthick %
         \global\advance\countREGISTER by 1%
         \ncase=0%
      \fi%
      \ifx\next\crnorule %
         \global\advance\countREGISTER by 1%
         \ncase=0%
      \fi%
      \ifx\next\crthickneg %
         \global\advance\countREGISTER by 1%
         \ncase=0%
      \fi%
      \ifx\next\crnoruleneg %
         \global\advance\countREGISTER by 1%
         \ncase=0%
      \fi%
      \ifx\next\crneg %
         \global\advance\countREGISTER by 1%
         \ncase=0%
      \fi%
      \ifx\next\headertab %
%     \out{In subROWcount:  next=header, ncase set=1}%
         \ncase=1%
      \fi%
%     \out{In subROWcount:  ncase is [\the\ncase]}%
      \relax%
      \ifcase\ncase %
         \let\next\ROWcount%
%        \out{subROWcount---> ncase=\the\ncase}%
      \or %
         \let\next\argROWskip%
%        \out{subROWcount---> ncase=\the\ncase}%
      \else %
      \fi%
   \fi%
%  \out{subROWcount---> NEXT=\meaning\next}%
   \next%
}%  End of macro \subROWcount.
\def\counthdROWS#1\into#2{%
\dvr{10}%
   \let\countREGISTER=#2%
   \countREGISTER=0%
\dvr{11}%
%  \out{In counthdROWS:  tokens are [\the#1]}%
\dvr{13}%
   \expandafter\hdROWcount\the#1\endcount%
\dvr{12}%
}%
\def\hdROWcount{%
   \afterassignment\subhdROWcount\let\next= %
}%
\def\subhdROWcount{%
%\out{In subhdROWcount:  next is [\meaning\next]}%
   \ifx\next\endcount %
      \let\next=\relax%
   \else%
      \ncase=0%
      \ifx\next\cr %
         \global\advance\countREGISTER by 1%
         \ncase=0%
      \fi%
      \ifx\next\endrow %
         \global\advance\countREGISTER by 1%
         \ncase=0%
      \fi%
      \ifx\next\crthick %
         \global\advance\countREGISTER by 1%
         \ncase=0%
      \fi%
      \ifx\next\crnorule %
         \global\advance\countREGISTER by 1%
         \ncase=0%
      \fi%
      \ifx\next\headertab %
%\out{In subhdROWcount:  next=header, ncase set=1}%
         \ncase=1%
      \fi%
%\out{In subhdROWcount:  ncase is [\the\ncase]}%
\relax%
      \ifcase\ncase %
         \let\next\hdROWcount%
%\out{subhdROWcount---> ncase=\the\ncase}%
      \or%
         \let\next\arghdROWskip%
%\out{subhdROWcount---> ncase=\the\ncase}%
      \else %
      \fi%
   \fi%
%\out{subhdROWcount---> NEXT=\meaning\next}%
   \next%
}%
{\catcode`\|=13\letbartab
\gdef\countCOLS#1\into#2{%
%  \out{In countCOLS:  tokens are [\the#1]}
   \let\countREGISTER=#2%
   \global\countREGISTER=0%
   \global\multispancount=0%
   \global\firstrowtrue
   \expandafter\COLcount\the#1\endcount%
   \global\advance\countREGISTER by 3%
   \global\advance\countREGISTER by -\multispancount
%  \out{countCOLS-->[\the\countREGISTER]}
}%
\gdef\COLcount{%
   \afterassignment\subCOLcount\let\next= %
}%
{\catcode`\&=13%
\gdef\subCOLcount{%
%\out{In subCOLcount: next is [\meaning\next]}
   \ifx\next\endcount %
      \let\next=\relax%
   \else%
      \ncase=0%
      \iffirstrow
         \ifx\next& %
            \global\advance\countREGISTER by 2%
            \ncase=0%
         \fi%
         \ifx\next\span %
            \global\advance\countREGISTER by 1%
            \ncase=0%
         \fi%
         \ifx\next| %
            \global\advance\countREGISTER by 2%
            \ncase=0%
         \fi
         \ifx\next\|
            \global\advance\countREGISTER by 2%
            \ncase=0%
         \fi
         \ifx\next\multispan
            \ncase=1%
            \global\advance\multispancount by 1%
         \fi
         \ifx\next\headertab
            \ncase=2%
         \fi
         \ifx\next\cr       \global\firstrowfalse \fi
         \ifx\next\endrow   \global\firstrowfalse \fi
         \ifx\next\crthick  \global\firstrowfalse \fi
         \ifx\next\crnorule \global\firstrowfalse \fi
         \ifx\next\crnoruleneg \global\firstrowfalse \fi
         \ifx\next\crthickneg  \global\firstrowfalse \fi
         \ifx\next\crneg       \global\firstrowfalse \fi
      \fi%  End of \iffirstrow.
\relax%\out{subCOL-->  ncase=[\the\ncase]}
% \out{subCOL-->  next=\meaning\next}
      \ifcase\ncase %
         \let\next\COLcount%
      \or %
         \let\next\spancount%
      \or %
         \let\next\argCOLskip%
      \else %
      \fi %
   \fi%
%  \out{subCOL-->  countREGISTER=[\the\countREGISTER]}
   \next%
}%
\gdef\argROWskip#1{%
%  Deletes the next balanced, undelimited argument from a
%                 token list.
% \out{---> Entering argROWskip <---}
% \out{In argROWskip:  deleted arg is [#1]}%
   \let\next\ROWcount \next%
}%  End of macro \argskip.
\gdef\arghdROWskip#1{%
%  Deletes the next balanced, undelimited argument from a
%                 token list.
% \out{---> Entering arghdROWskip <---}
% \out{In arghdROWskip:  deleted arg is [#1]}%
   \let\next\ROWcount \next%
}%  End of macro \arghdROWskip.
\gdef\argCOLskip#1{%
%  Deletes the next balanced, undelimited argument from a
%                 token list.
% \out{---> Entering argCOLskip <---}
% \out{In argCOLskip:  deleted arg is [#1]}%
   \let\next\COLcount \next%
}%  End of macro \argskip.
}%  End of active &'s.
}%  End of active |'s.
\def\spancount#1{%\out{spancount--->\meaning#1}
   \nspan=#1\multiply\nspan by 2\advance\nspan by -1%
   \global\advance \countREGISTER by \nspan
%  \out{number spancount--->\the\nspan; \the\countREGISTER}
   \let\next\COLcount \next}%
\def\dvr#1{\relax}%
% \omit\hfil%
% \parindent=0pt\hsize=1.1in\valign{%
% \vfil#\vfil&\vfil#\vfil\cr\hfil\hbox{\ Added to\ }\hfil&%
% \hfil\hbox{\ empty events\ }\hfil\cr}\hfil%
\def\headertab#1{%
\dvr{1}{\let\cr=\@mpersand%
\hdtks={#1}%
%\out{In header:  hdtks=[\the\hdtks]}%
\counthdROWS\hdtks\into\hdrows%
\advance\hdrows by 1%
\ifnum\hdrows=0 \hdrows=1 \fi%
%\out{In header:  Nhdrows=[\the\hdrows]}%
\dvr{5}\makehdPREAMBLE{\the\hdrows}%
%\out{In header:  headerpreamble=[\headerpreamble]}%
\dvr{6}\getHDdimen{#1}%
%\out{In header:  hdsize=[\the\hdsize]}%
%\striplastCR{#1}%
{\parindent=0pt\hsize=\hdsize{\let\ifmath0%
\xdef\next{\valign{\headerpreamble #1\crnorm}}}\dvr{7}\next\dvr{8}%
}%
}\dvr{2}}%  End of macro \headertab.
\def\makehdPREAMBLE#1{%This macro generates the necessary preamble for a
\dvr{3}%
%                      ruled table with \ncols primary columns.
%                      (Primary columns means the number of columns NOT
%                       counting those used for vertical rules.
\hdrows=#1%  Get the number of columns desired.
{%  Start local parameter definitions.
\let\headerARGS=0%
%  This is the key to the whole thing; it prevents \ARGS
\let\cr=\crnorm%
%                from being expanded in the followin \edef's.
\edef\xtp{\vfil\hfil\hbox{\headerARGS}\hfil\vfil}%
\advance\hdrows by -1%  One row has been generated; decrement the
%                         counter.
\loop%  Append as many further rows as needed to the preamble.
\ifnum\hdrows>0%
\advance\hdrows by -1%
\edef\xtp{\xtp&\vfil\hfil\hbox{\headerARGS}\hfil\vfil}%
\repeat%
\xdef\headerpreamble{\xtp\crcr}%
}%  End of local parameters.
\dvr{4}}%  End of \makehdPREAMBLE.
\def\getHDdimen#1{%
%\out{In getHDdimen:  Arg 1=[#1]}%
\hdsize=0pt%
\getsize#1\cr\end\cr%
}%  End of macro getHDdimen.
\def\getsize#1\cr{%
%\out{In getsize:  Arg 1=[#1]}%
%  Here we have to check arg#1 and see if the first token in #1 is an
%    \end; if so, we stop, else we check the width of arg#1.
%  We recall that each arg#1 will be terminated with a \cr token.
\endsizefalse\savetks={#1}%
%\out{In getsize:  the savetks = [\the\savetks]}%
\expandafter\lookend\the\savetks\cr%
%\out{In getsize:  ifendsize = [\meaning\ifendsize]}%
\relax \ifendsize \let\next\relax \else%
\setbox\hdbox=\hbox{#1}\newhdsize=1.0\wd\hdbox%
\ifdim\newhdsize>\hdsize \hdsize=\newhdsize \fi%
%\out{In getsize:  hdsize=[\the\hdsize]}%
%\out{In getsize:  newhdsize=[\the\newhdsize]}%
\let\next\getsize \fi%
\next%
}%
\def\lookend{\afterassignment\sublookend\let\looknext= }%
\def\sublookend{\relax%
%\out{In sublookend:  looknext = [\looknext]}%
\ifx\looknext\cr %
%\out{In sublooknext:  looknext=cr}%
\let\looknext\relax \else %
%\out{In sublooknext:  looknext/=cr}%
   \relax
   \ifx\looknext\end \global\endsizetrue \fi%
   \let\looknext=\lookend%
    \fi \looknext%
}%
%
%  Allow the user to make his own names for crthick, etc.
%
\def\tablelet#1{%
   \tableLETtokens=\expandafter{\the\tableLETtokens #1}%
}%
\catcode`\@=12%  Change @'s back to their normal category code.

\def\vs{\noalign{\vskip5pt}}
\def\rmind{{\rm ind}}
\def\rmdep{{\rm dep}}
\def\Q{Q_{\rm rms-ps}}
\def\qup{\Q^{\rm upper\ limit}}
\def\qlow{\Q^{\rm lower\ limit}}
\def\nch{N_{\rm ch}}
\def\npat{N_{\rm p}}
\def\rind{{\rm ind}}
\def\ct{C^{\rm mbr}}
%%%%%%%%%%%%%%%%%%%%%%%%%%%%%%%%%%%%%%%%%%%%%%%%%%%%%%
%%%%%%%%%%%% Equation Numbering %%%%%%%%%%%%%%%%%%%%%%
%%%%%%%%%%%%%%%%%%%%%%%%%%%%%%%%%%%%%%%%%%%%%%%%%%%%%%
\def\be#1{$$\global\advance\counteqn by 1 \xdef#1{(\number\counteqn)}}
\def\ee{\eqno(\number\counteqn)$$}
\def\ben{$$\global\advance\counteqn by 1}
\def\bea#1#2{\global\advance\counteqn by 1 \xdef#1{(\number\counteqn)}
             $$\eqalignno{#2 &(\number\counteqn)\cr}$$}
\def\bean#1{\global\advance\counteqn by 1 
            $$\eqalignno{#1 &(\number\counteqn)\cr}$$}
%%%%%%%%%%%%%%%%%%%%%%%%%%%%%%%%%%%%%%%%%%%%%%%%%%%%%%
%%%%%%%%%%%%      Figures       %%%%%%%%%%%%%%%%%%%%%%
%%%%%%%%%%%%%%%%%%%%%%%%%%%%%%%%%%%%%%%%%%%%%%%%%%%%%%
\def\putfig#1#2#3{
\goodbreak\midinsert
\global\advance\countfig by 1
\xdef#1{\number\countfig}
\centerline{\epsfysize=5truein\epsfbox{#2}}
{\figfont \noindent FIG.\ \number\countfig. #3}
\endinsert}

\def\putfigp#1#2#3{
\global\advance\countfig by 1
\xdef#1{\number\countfig}
{\figfont \noindent FIG.\ \number\countfig. #3}
}

\font\largerm=cmr10  at 12.00pt
\font\largebf=cmbx10 at 12.00pt
\font\figfont=cmr8

\newcount\countsec
\newcount\counteqn
\newcount\countfig
{\nopagenumbers
\vskip15cm
\centerline{\largebf ANALYSIS OF SMALL SCALE MBR ANISOTROPY}
\centerline{  {\largebf IN THE PRESENCE OF FOREGROUND CONTAMINATION}
\footnote{$^1$}{
Submitted to {\it The Astrophysical Journal}}
}
\bigskip
\bigskip
\centerline{\largerm SCOTT DODELSON and ALBERT STEBBINS}
\bigskip
\bigskip
\centerline{\largerm NASA/Fermilab Astrophysics Center}
\centerline{\largerm Fermi National Accelerator Laboratory}
\centerline{\largerm P.O. Box 500, Batavia, IL 60510      }
\centerline{\largerm USA}
\centerline{\sl dodelson@virgo.fnal.gov}
\centerline{\sl stebbins@fnalv.fnal.gov}
\bigskip
\bigskip
\bigskip
\bigskip

\centerline{\largebf Abstract}

Many of the current round of experiments searching for anisotropies in the
Microwave Background Radiation (MBR) are confronting the problem of how to
disentangle the cosmic signal from contamination due to galactic and
intergalactic foreground sources. Here we show how commonly used likelihood
function techniques can be generalized to account for foreground.  Specifically
we set some restrictions on the spectrum of foreground contamination but allow
the amplitude to vary arbitrarily.  The likelihood function thus generalized
gives reasonable limits on the MBR anisotropy which, in some cases, are not
much less restrictive than what one would get from more detailed modeling of
the foreground.  Furthermore, the likelihood function is exactly the same as
one would obtain by simply projecting out foreground contamination and just
looking at the reduced data set.  We apply this generalized analysis to the
recent medium angle data sets of ACME-HEMT (Gaier~\etal 1992, Schuster~\etal
1993) and MAX (Meinhold~\etal 1993, Gunderson~\etal 1993).  The resulting
analysis constrains the one free parameter in the standard cold dark matter
theory to be $\Q=18_{-5}^{+8}\mu K$. This best fit value, although in striking
agreement with the normalization from COBE, is not a very good fit, with an
overall $\chi^2/$ degree of freedom $=208/168$. We also argue against three
commonly used methods of dealing with foreground: (i) ignoring it completely;
(ii) subtracting off a best fit foreground and treating the residuals as if
uncontaminated; and (iii) culling data which appears to be contaminated by
foreground.
%\footnote{}{
%{\it Subject headings:} cosmology: cosmic background radiation }

\vfill\eject}

\global\advance\countsec by 1
\pageno=1

\header{\number\countsec. Introduction}

	Since the detection of Microwave Background Radiation (MBR)
anisotropies by the Differential Microwave Radiometer (DMR) on the COsmic
Background Explorer (COBE) satellite (Smoot \etal 1992), there has been a spate
of announcements of detections of anisotropies in microwave brightness on
smaller angular scales (Gaier~\etal 1993, Meinhold~\etal 1993, Schuster~\etal
1993, Gunderson~\etal 1993, and more).  Many of these experiments utilize
measurements at multiple frequencies in order to be able distinguish primordial
anisotropies in the MBR from other types of microwave emission, such as dust,
free-free, or synchrotron, which may occur either in our Galaxy or in
extra-Galactic objects.  While the large scale anisotropies observed by the
COBE-DMR appear to have relatively little foreground contamination, the
spectrum of many of the small scale anisotropies found are not very well fit by
frequency-independent brightness temperature fluctuations and are therefore
probably not completely due to primordial anisotropy.  As stressed by Brandt
\etal, (1993) it will take more and better observations to be able to
disentangle the primordial anisotropies from the foreground contamination.  In
the meantime one would like to use the small-scale measurements to set limits
on the parameters of models of
cosmological inhomogeneities. To obtain such limits one must take into account
the uncertain contamination of the measured anisotropies.  In this paper we
will discuss some methods of accounting for the contamination when constraining
parameters and apply them to some of the ACME-HEMT (Gaier~\etal 1992,
Schuster~\etal 1993) and MAX (Meinhold~\etal 1993, Gunderson~\etal 1993) data.
We stress that taking into account unknown amounts of contamination
involves great uncertainties and different approaches can be expected to yield
different limits on parameters.  Here the emphasis will be on setting reliable,
and perhaps conservative, limits on model parameters.

	In \S2 we introduce the likelihood function which is the standard tool
used to compare theoretical predictions with MBR anisotropy data.  We then go
on to discuss various properties of the generalization of the likelihood
function to include a known statistical distribution of sources.  We show that
this generalized likelihood function has a well-defined and useful limit when
we take the amplitude of the foreground anisotropies to be large.  This limit
is independent of the spatial correlations of the foreground emission and
depends only on the assumed spectra of the various components of foreground
contamination.  This limit is also equivalent to ``projecting out'' or
``marginalizing'' (Anthony Lasenby's terminology) the foreground emission
from multifrequency data.  Finally in \S2 we show how this limit may be easily
taken when the MBR model and detector noise are assumed Gaussian.  In \S3 we
give some examples of how the procedure behaves by applying it to the data of
Gaier~\etal (1993).  In \S4 we apply it to rest of the data.  Finally in \S5 we
discuss the meaning of the results obtained.

\global\advance\countsec by 1

\header{\number\countsec. Likelihood Functions}

	There are at least three components to microwave temperature
anisotropies measured on the sky.  For cosmologists the most interesting
component is the set of
primordial MBR anisotropies.  These are caused by the initial
inhomogeneities in the Universe which are thought to be set up by some random
process occurring at very early cosmological times.  The particular random
process we will refer to as a cosmological model.  Here we will suppose one is
considering only a subset of possible models, parameterized by a few numbers.
Let us denote these numbers by the shorthand $\calP$, for parameters.  What we
would like to do is estimate $\calP$ from a set of MBR anisotropy measurements.
Let $\calT$ represent the {\it true} values of the primordial MBR anisotropies
that have been estimated by experiments.  Given $\calT$ we can estimate
$\calP$ using the probability density of $\calT$ given $\calP$, i.e.
$p_{\s1}(\calT|\calP)\,d\calT$.  In addition to the primordial MBR anisotropy
there is also a component of foreground contamination which adds to the signal
coming into an MBR experiment, or symbolically $\calS=\calT+\calF$, where
$\calS$ and $\calF$ refer to external signal and foreground contamination.
The third ingredient is the observational error or detector noise, $\cal N$, in
measuring brightness fluctuations.  Thus the data we obtain, $\calD$, can be
written as $\calD=\calS+\calN=\calT+\calF+\calN$. By testing and good
experimental design we usually know the distribution of $\calN$ or
mathematically $p_{\s2}(\calD|\calS)\,d\calD$ which is the probability density
of $\calD$ given $\calS$.  Unfortunately we do not know much about the
properties of $\calF$ and this limits what we can say about $\calP$.

	For the moment assume $\calF$ is zero, as is done in many analyses of
MBR experiments.  Then one has all of the ingredients to construct the
probability density of $\calD$ conditional on $\calT$, i.e.
\be\PROBDAT
p_{\s3}(\calD|\calP)d\calD=
\left[\int p_{\s2}(\calD|\calT)\ p_{\s1}(\calT|\calP)\,d\calT\right]\,d\calD\ .
\ee
One approach to inferring $\calP$ from $\calD$ is to choose $\calP$ such that
for that value of $\calP$ the observed data, $\calD$, is near [but not too
near] the mean of the distribution.  Along the same lines one might require
that $\calP$ be chosen such that $\calD$ has a high probability density, when
compared with other possible values of $\calD$.  These methods may be described
as requiring ``goodness-of-fit'' of the data and involve fixing $\calP$ and
looking at the probability distribution of $\calD$.  Of course what we would
really like is the probability distribution of $\calP$.  Unfortunately there is
nothing in probability theory that would allow us to convert the data into such
a probability distribution without further assumptions.

	Another way to infer $\cal P$ from $\calD$ involves not the
probability of $\calP$, but its likelihood.  That is, fix the data at the
observed value, $\calD$, and choose $\calP$ such that the probability density,
$p_{\s3}(\calD|\calP)$, be large when compared to other values of $\calP$.
Clearly a large probability density near the observed data favors such a value
of $\calP$.  Another motivation for this procedure stems from Bayes' Theorem.
Let us suppose we had made the further assumption (or had further knowledge)
that the parameters, $\calP$, were determined by some random process, with
probability distribution $p_{\s4}(\calP)\,d\calP$.  We wish to evaluate the
$\calP$ used to generate the data, $\calD$.  Bayes' theorem says that given
that the measurements yield $\calD$, the new ({\it posterior}) probability
distribution for $\calP$ is just the original ({\it prior}) distribution times
the likelihood function, $L$, or mathematically
\be\BAYES
 p_{\s5}(\calP|\calD)\,d\calP=L(\calP;\calD)\,p_{\s4}(\calP)\,d\calP \qquad
L(\calP;\calD)\equiv
{p_{\s3}(\calD|\calP)\over\int p_{\s3}(\calD|\calP)\,p_{\s4}(\calP)\,d\calP}\ .
\ee
Note that the ratio of $L$ for different values of $\calP$, which is really
just the ratio of $p_{\s3}(\calD|\calP)$, does not depend on the prior
distribution.  It is this likelihood ratio which gives the relative increase or
decrease of the probability of different values of $\calP$ and this is why one
might consider values of $\calP$ with relatively high likelihood as being
preferred.  Even if one doesn't have any knowledge of the prior distribution of
$\calP$ one can still use the likelihood ratio as a statistic to choose
$\calP$, although one then cannot determine the probability of the favored
choice being correct.

	In recent years the likelihood function has been commonly used as a
statistic in analyzing anisotropy experiments 
(see Readhead \etal (1989) or Myers (1990) for clear discussions), 
although usually without
regard to any possible foreground contamination.  If contamination is present
then one must make some assumptions about the contamination in order to
construct the likelihood function.  If one knows or guesses that the foreground
is drawn from some probability distribution, $p_{\s6}(\calF)\,d\calF$, then
\be\LIKE
L(\calP;\calD)\propto p_{\s3}(\calD|\calP)
=\int\int p_{\s2}(\calD|\calT+\calF)\ p_{\s1}(\calT|\calP)
                                    \ p_{\s6}(\calF)\,d\calT\,d\calF
\ee
where we have dropped the $\calP$-independent denominator of $L$.  One could
use one's general knowledge of the Galactic dust and gas and of extra-Galactic
sources to construct a reasonable $p_{\s6}$.  If available one should make use
of measurements of the emission at other frequencies to constrain the
foreground emission in the same direction one is measuring the microwave
brightness anisotropy.   Some radio and infrared data are available for 
the measurements considered below and we will give some discussion of these in
\S5.  For the moment we will consider the case where we do not have any 
compelling reason to chose one $p_{\s6}(\calF)$ over another.  We now discuss
how one can still set believable limits on model parameters even in this state
of ignorance.

\subheader{\bf Marginalization}

	If we really do not have much idea of what the foreground is doing then
to set reliable (= conservative) limits on MBR anisotropy we should take a
liberal view of what the foreground emission may do to contaminate the
measurements. One might think that such an approach will lead to no limits at
all, but this is not the case.  If one sets some restrictions on the spectrum
of foreground contamination but allows the amplitude to vary arbitrarily then
the likelihood function gives reasonable limits on the MBR anisotropy which, in
some cases, are not much less restrictive than what one would get from more
detailed modeling of the foreground.  Furthermore, in this limit the likelihood
function is exactly the same as one would obtain by simply projecting out
foreground contamination and just looking at the reduced data set.

	Before proceeding to justify these claims we should be a bit more
explicit about the meaning of $\calD$, $\calT$, and $\calF$.  The brightness
pattern on the sky, $I_\nu(\hatbfn)$, is a continuous function of both
direction, $\hatbfn$, and frequency, $\nu$.  It is the $\nu$ dependence which
we can use to unambiguously separate MBR anisotropy from foreground
contamination.  We therefore only consider multifrequency experiments. The
experiments only give estimates of the temperature convolved with some window
function in direction and frequency, i.e.  
\be\SIGNAL
\calS =\left\{\int W_i(\hatbfn)\,
V_{\s(i,a)}(\nu)\,I_\nu(\hatbfn)\,d^2\hatbfn\,d\nu ,\ 
              a=1,\ldots,N_{\rm ch},\,i=1,\ldots,N_\rmp\right\}          \ .
\ee
where $N_\rmp$ gives the number of spatial {\it patch}es and $N_{\rm ch}$ gives
the number of spectral {\it channel}s which for simplicity we have assumed is
the same for all patches.  For simplicity we have also assumed that the window
functions factorize into spatial $(W_i(\hatbfn))$
and spectral $(V_{\s(i,a)}(\nu))$
parts and that the spatial window
is the same for each channel at that patch.  In general we only
require that the window function averaged over frequency and weighted by the
spectra of any component, i.e. MBR or foreground, be the same for all
components and all channels.  With this assumption the relation between the
signal in different channels of the same patch is telling us only about the
spectrum and the detector noise, and not about the spatial pattern of
brightness.  The moments of both the MBR and foreground brightness which
contribute to the measured signal are just the same convolution as given in
eq.~\SIGNAL.  The MBR brightness in a given direction is given by only one
parameter, the temperature $T_{\s\rm MBR}(\hatbfn)$.  We will assume that there
are only a finite number of components of contamination, say $N_\rmf$, each
of which have a {\it known} spectrum.  Furthermore we assume that
$N_\rmf<N_{\rm ch}$.  If not then any possible observed signal
could always be produced by some combination of foreground emission and no MBR
anisotropy.  Finally $\calD$ is just $\calS$ of eq.~\SIGNAL\ with some added
detector noise which is different in each patch and channel.

	Let us count the number of degrees of freedom (dof) of the various
terms which contribute to the $N_\rmp\times N_{\rm ch}$ observations, $\calD$:
$\calT$ has $N_\rmp$ dof, $\calF$ has $N_\rmp\times N_\rmf$ dof, and
$\calN$ has $N_\rmp\times N_{\rm ch}$ dof.  This means that there are $(N_{\rm
ch}-N_\rmf)\times N_\rmp$ dof of $\calD$ which are linearly independent of
$\calF$.  Thus even if we let $\calF$ span its entire range the resultant
$\calD$ does not span the entire space of observations, and this why a liberal
attitude toward $\calF$ still yields interesting results.

	To implement our liberal approach toward $\calF$ consider the class of
probability distributions for $\calF$ 
\ben
p_{\s6}(\calF)d\calF=\alpha\,f_{\s6}(\alpha \calF)d\calF \ .
\ee
Any normalizable distribution must fall off for large $\calF$, but as
$\alpha\rightarrow0$ the region of significant probability density gets larger
and larger and the variance of this prior distribution increases.  However the
likelihood function of eq.~\LIKE\ will remain well behaved since for fixed 
$\calD$,
$p_{\s2}(\calD|\calT+\calF)$ will fall off.  Note that $\calT$ and $\calF$
cannot exactly
cancel since they have different spectra.  Thus in the limit of large
variance we may replace $p_{\s6}(\calF)$ with $p_{\s6}(0)$, and since this
constant is independent of $\calP$ it will not enter into any likelihood ratio
and we may drop it.  Of course we require that  $p_{\s6}(0)\ne0$, however this
is true of any reasonable distribution. For large variance we have
\be\LIKEMARG
L(\calP;\calD,\calF)\rightarrow L^*(\calP;\calD)\propto
\int\int p_{\s2}(\calD|\calT+\calF)\ p_{\s1}(\calT|\calP)\,d\calT\,d\calF \ .
\ee
Thus we see that this likelihood function is just what we would get for a
uniform prior for $\calF$.  Such a uniform prior may include correlations
between the different patches or correlations between different components of a
multi-component foreground and the determinant of this correlation matrix may
enter into $p_{\s6}(0)$, however this will not effect the likelihood ratio of
eq.~\LIKEMARG. Thus we are lead to our first interesting result:

\item{$\bullet$} In the limit of large variance all prior distributions for the
foreground contamination yield the same likelihood function.

\noindent We may simplify Eq.~\LIKEMARG\ still further if we make the
assumption that the probability distribution of detector noise does not depend
on the amplitude of the signal, i.e.
\be\NOISE
p_{\s2}(\calD|\calT+\calF)=f_{\s2}(\calD-\calT-\calF) \ .
\ee
To see how this helps let us denote $\calD$'s dof which are independent of
$\calF$ by $\calD^\rmind$ and the dependent dof by $\calD^\rmdep$. The MBR
contribution to the signal may be similarly decomposed, but the foreground of
course has no independent part.  We may thus rewrite Eq.~\NOISE\ as 
\ben
p_{\s2}(\calD|\calT+\calF)
         =f_{\s2}(\calD^\rmind-\calT^\rmind,\calD^\rmdep-\calT^\rmdep-\calF)\ .
\ee
If we substitute this into eq.~\LIKEMARG\ and change one of the variables of
integration from $\calF$ to $\calD^\rmdep$ we find
\be\LIKEMARGIND
L^*(\calP;\calD)\propto
\int\left[\int p_{\s2}(\calD|\calT+\calF)d\calD^\rmdep\right]
                                               \,p_{\s1}(\calT|\calP)\,d\calT
                \propto L(\calP;\calD^\rmind) \ .
\ee
since the term in square brackets is just the marginal distribution of
$\calD^\rmind$ obtained by integrating out $\calD^\rmdep$. This gives us our
second interesting result

\item{$\bullet$} The likelihood function constructed by using all of the
data and a prior distribution for the foreground contamination with very large
variance is equal to the likelihood function obtained by considering only that 
linear combination of the data which is independent of the foreground
contamination.

\noindent One can obtain $\calD^\rmind$ by projecting out all linear
combinations of the spectrum of the different foreground components. This last
identity is of great practical use since it reduces the dimensionality of the
data space and can make computations of the likelihood function much more
tractable.  Of course this idea of projecting out the foreground contamination
is not a new one.  For example, it has been used in constructing the ``reduced
galaxy'' (RG) anisotropy map of the DMR experiment.  However the above equality
does give an added justification for constructing such RG data sets, as it
shows that by doing so one is not throwing out any information except for ones
assumptions about the properties of the foreground contamination.

	The likelihood function $L^*$ is obtained by integrating over
arbitrarily large foreground contamination, which is certainly not true.  Given
some true distribution of foreground contamination, $p_{\s6}(\calF)$, under
what circumstances will $L^*$ give a good approximation to the true likelihood
function?  As mentioned before, $p_{\s2}$ in Eq.~\LIKE\ will regulate the
integral over $\calF$ if $p_{\s6}$ does not.  Since $p_{\s2}$ will start
falling roughly when the foreground emission exceeds either the detector noise
or the observed signal it follows that the condition for $L^*$ to be an
accurate representation of $L$ is roughly that the variance in the foreground
emission exceeds greatly either the detector noise or the observed signal.

	In obtaining these results we have made no assumptions about the
statistical properties of the MBR anisotropy or the foreground contamination,
although one will have to make some assumptions about the former in order to
compute $L^*$.  In particular we haven't assumed anything is Gaussian.  We have
assumed that the detector noise is independent of the amplitude of the signal,
however we think this likely to be a very good approximation.  More important
assumptions were made about the experimental apparatus, in particular we have
assumed that the effective window function is the same for all channels of a
given patch.  For some experiments, such as the DMR, this is an excellent
approximation, while for other such as the ACME experiment it is not. The
degree to which this varying window can confuse spatial and spectral
dependencies will depend on the spatial distributions of the emission.  One can
alleviate this problem if the region of the sky is oversampled by the
experiment, in which case one can bin the data into synthetic beam patterns
which are the same for all channels.  Finally we have assumed that we know the
spectrum of the various foreground contaminants.  Here it is important
only that deviations from the assumed spectra are not sufficient that the
$\calD^\rmind$ used in the likelihood function could be significantly
contaminated by foreground emission.  Here we are making some assumption about
the amplitude of foreground emission, but if our assumed spectrum is fairly
accurate then this is a much less stringent constraint than assuming that the
foreground emission is small compared with the observed signal.  In the
microwave region there is fairly small uncertainty in the spectra of free-free
emission, however for synchrotron and dust emission this is not the case.
Hopefully data from the DIRBE and FIRAS experiments on COBE will show a fairly
universal spectra for dust emission at microwave frequencies.  We do know that
at longer wavelengths synchrotron emission exhibits an unfortunately broad
range of spectral indices and this can lead to large uncertainty in
constructing $\calD^\rmind$.

\subheader{Gaussian Statistics}

	Now we can apply this analysis specifically to the case where the MBR
anisotropies are Gaussian random noise with some unknown parameters and the
detector noise is also Gaussian.  Let us represent the data, $\calD$, by a set
of numbers $\Delta_{\s(a,i)}$ which give the signal in channel $a$ of patch 
$i$.  We may similarly represent the contribution of the foreground
contamination to each such patch and channel as  $\Delta^\rmf$.  Thus the MBR
anisotropy plus detector noise, $\calT+\calN$, is given by
$\Delta_{\s(a,i)}-\Delta^\rmf_{\s(a,i)}$, which has the correlation matrix
\ben
\left\langle(\Delta_{\s(a,i)}-\Delta^\rmf_{\s(a,i)})
              (\Delta_{\s(b,j)}-\Delta^\rmf_{\s(b,j)})\right\rangle=
                                              C_{\s(a,i)(b,j)}  \qquad
C_{\s(a,i)(b,j)}=C^{\rm mbr}_{ij}
                   +\sigma^2_{\s(a,i)}\delta_{ij}\delta_{ab}
\ee
where $\sigma_{\s(a,i)}$ gives the instrumental noise and $C^{\rm mbr}_{ij}$
gives the expected correlation in MBR fluctuations.    We will assume that
we have some model which determines $C^{\rm mbr}_{ij}$ modulo the value of some
parameters which are the $\calP$ of the previous discussion.  The MBR
anisotropy contribution to the signal and the instrumental noise  are assumed
Gaussianly distributed with zero mean and are thus fully determined by their
respective correlation matrices, $C^{\rm mbr}_{ij}$ and
$\sigma^2_{\s(a,i)}\delta_{ij}\delta_{ab}$.  The foreground emission may be
written as a sum over the different foreground emission processes (e.g.
free-free, synchrotron, dust), i.e.  
\ben
\Delta^\rmf_{\s(a,i)}
                   =\sum_{\alpha=1}^{N_\rmf} A_{\s[\alpha,i]}F_{\s\{\alpha,a\}}
\ee
where $\alpha$ labels the process, $A_{\s[\alpha,i]}$ gives the amplitude of
emission of process $\alpha$ in beam $i$, and $F_{\s\{\alpha,a\}}$ gives the
contribution per unit amplitude of process $\alpha$ in channel $a$.  The
quantities $F_{\s\{\alpha,a\}}$ are assumed known, however the amplitudes
$A_{\s(\alpha,i)}$ are not.  It is these $A_{\s[\alpha,i]}$'s which represent
the $\calF$ in the previous subsection.

	One can explicitly compute the components of $\Delta_{\s(a,i)}$ which
are independent of the foreground by solving the system of equations
\be\NULLSPACE
\sum_{a=1}^{N_{\rm ch}}z_a F_{\s\{\alpha,a\}}=0 \qquad\alpha=1,\ldots,N_\rmf,
\ee
say with a general purpose eigensystem solver.  As long as one is able to
distinguish the different components of foreground contamination, i.e. as long
as $F_{\s\{\alpha,a\}}$ are linearly independent, there will be an $(N_{\rm
ch}-N_\rmf)$-dimensional space of solutions of eq.~\NULLSPACE.  If one chooses
a basis for this space, $z^{\s(r)}_a$, where $r$ labels the basis
vectors, then one obtains a set of coordinates on the data subspace which are
independent of the foreground:
\ben
\Delta^\rmind_{\s|r,i|}
                 =\sum_{a=1}^{N_{\rm ch}}z^{\s(r)}_a\Delta^\rmind_{\s(a,i)} \ .
\ee
The correlation function on $\calD^\rmind$ is 
\be\CORRIND
\left\langle \Delta^\rmind_{\s|r,i|}\Delta^\rmind_{\s|s,j|}\right\rangle
        =C^\rmind_{\s|r,i||s,j|}=\sum_{a=1}^{N_{\rm ch}}\sum_{b=1}^{N_{\rm ch}}
                                        z^{\s(r)}_a C_{\s(a,i)(b,j)}z^{\s(s)}_b
\ee
so the likelihood functions of eq.~\LIKEMARGIND\ are
\ben
L^*(\calP;\calD)\propto L(\calP;\calD^\rmind)\propto 
{\exp\left(-{1\over2}\chi_\rmind^2\right)\over\Vert C_{\s|r,i||s,j|}\Vert}
\ee
where $\Vert\ \Vert$ indicates the determinant and  
\be\CHIIND
\chi_\rmind^2=\sum_{i=1}^{N_\rmp}           \sum_{j=1}^{N_\rmp}
              \sum_{r=1}^{N_{\rm ch}-N_\rmf}\sum_{s=1}^{N_{\rm ch}-N_\rmf}
           \Delta^\rmind_{\s|r,i|}C^{\rmind^{\scriptstyle-1}}_{\s|r,i||s,j|}
           \Delta^\rmind_{\s|s,j|} \ .
\ee
Of course, $\chi_\rmind^2$ is just the chi-square statistic which for the
correct choice of $C^\rmind_{\s|r,i||s,j|}$ will be distributed like a
$\chi^2$-distribution with $N_\rmp\times(N_{\rm ch}-N_\rmf)$ dof.

\global\advance\countsec by 1

\header{\number\countsec. Example: South Pole 91}

In this section, we will flesh out the formalism we've set up with an example.
We will restrict our analysis to testing one particular theory, cold dark
matter (CDM) with a Harrison- Zel'dovich initial 
spectrum.  We'll focus on an
experiment that's been analyzed a number of times already, so at least the
first part of our discussion, which assumes there are no foreground sources,
should be familiar to many.  This experiment, which we refer to as SP91, was
performed at the South Pole in 1991 with the ACME-HEMT telescope, and the
results were published in Gaier~\etal (1992).

\subheader{\bf One Patch; One Frequency Channel}

We start by considering the simplest possibility: An experiment measures the
temperature difference in one region of the sky at one frequency. This one
measurement, call it $\Delta$, is thus the full set of our data $\cal D$. The
first thing we need to know is what is the temperature distribution -- the full
set $\cal T$ is now simply one number $T$ -- predicted by the theory,
$p_1(T\vert\Q)$.  The one parameter in the theory [the set previously denoted
$\calP$] is the normalization, which we'll take as $\Q$, the average quadrupole
over an ensemble of universes. Since perturbations are Gaussian in standard
cold dark matter, we can write
\ben p_1(T | \Q) = {1\over \sqrt{2\pi \ct}}\ \exp
		\left\{ -{1\over2} {T^2\over \ct} \right\}
				.
\ee
The width of the Gaussian here, $\ct$,
depends on both the theory and the experimental configuration of the
beam. In particular,
\be\ELSUM
\ct = \sum_{l=2}^\infty {2l+1\over 4\pi} C_l W_l
\ee
where $C_l$ is the coefficient of the Legendre polynomial
$P_l(\hatn_1\cdot\hatn_2)$ when $C(\hatn_1\cdot\hatn_2)
\equiv\langle T(\hatn_1)T(\hatn_2)\rangle$ is expanded in a series of such
polynomials. That is, the $C_l$'s are given by the theory. The theory's one
free parameter, $\Q$, is related to $C_2$ via: $C_2=4\pi Q^2/5$. Meanwhile, the
window function $W_l$ is solely determined by the experimental beam size and
chopping strategy. For the SP91 experiment (Bond \etal 1991)
\be\EFIL
W_l = \exp\left\{-l(l+1)\theta_s^2 \right\}
	{16\pi\over 2l+1} \sum_{m=-l}^l H_0^2(m\phi_A) Y_{lm}^2(\theta_z,0)
\ee
where $\theta_z = 27.75^\circ$; $\phi_A = 1.5^\circ/\sin(\theta_z)$; $\theta_s
= 0.425 \times 1.35^\circ$ [for the highest frequency channel we are discussing
at present]; and $H_0$ is the Struve function of order zero. The $C_l$'s for
CDM and $W_l$ for SP91 are plotted in Figure 1.  Once the $C_l$ and $W_l$ are
given, it is straightforward to combine them and compute the expected variance
$\ct$. For standard CDM and SP91, we find $\ct = (43 \Q/17)^2$. [Recall that
COBE-normalized CDM has $\Q=17\mu$K.]

\putfig\wlcl{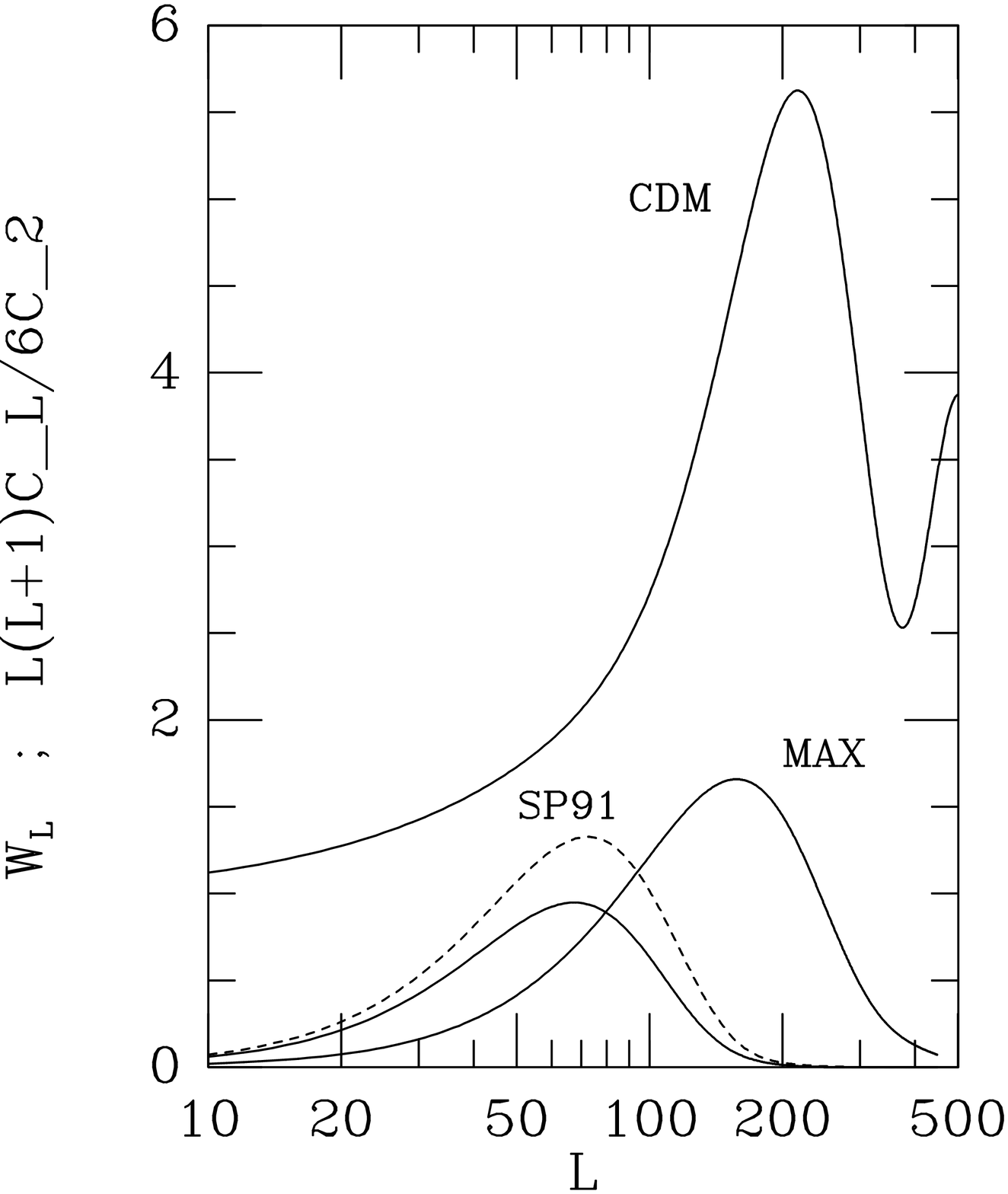}{The window function $W_l$ for the SP91 experiment and
the $C_l$'s in CDM with $h=0.5;\Omega_B=0.06;n=1$. Plotted is $l(l+1)C_l/6$
which is equal to one if only the Sachs-Wolfe effect is considered.  The dashed
line is an approximation to the SP91 filter function [assuming square well
chopping] which is seen to be off by as much as $30\%$. Also shown is the
filter function for the MAX experiments to be analyzed in the next section.}

The next step is to account for experimental errors by calculating
$p_2(\Delta \vert T)$. We assume the errors are Gaussian, so that
\be\EPTWO
p_2(\Delta \vert T) = {1\over\sqrt{2\pi}\,\sigma_{\rm exp}}\
			\exp\left\{-{1\over2}{(\Delta - T)^2
			           \over\sigma_{\rm exp}^2}\right\}.
\ee
Here, $\sigma_{\rm exp}$ is the variance of measurements in the `lab' where
there are no other sources [cosmic or otherwise] contributing to the signal.
So, eq.~\EPTWO\ simply tells us that if $\sigma_{\rm exp}$ were very small, the
observed value $\Delta$ would be very close to the actual value on the sky,
$T$. On the other hand, if the noise is significant, the observed value could
differ significantly from the sky value. One of the exciting things about the
SP91 experiment is that $\sigma_{\rm exp}$ was of order $20-30\mu$K,
significantly below the signal predicted by CDM.

Now that $p_1$ and $p_2$ are given, we can convolve the two as required
by eq.~\ELSUM\ to form $p_3(\Delta \vert \Q)$. The integral over $T$ is
readily performed and we find
\be
\EPTHREE p_3(\Delta \vert \Q) = {1\over (2\pi)^{N/2}}  \vert\vert C
\vert\vert^{-1/2}
		\exp\left\{ -{1\over2}  \Delta C^{-1} \Delta
				\right\} \ee
where $N$ is the number of measurements [here just one] and the correlation
``matrix'' [in this case one by one]
\ben C \equiv \langle \vert T + {\cal N} \vert^2 \rangle
		= \ct + \sigma_{\rm exp}^2 .
\ee
Figure 2 shows $p_3$ as a function of the observed temperature $\Delta$
for several values of the normalization $\Q$.
Here is a good time to hearken back to our discussion following eq.~\PROBDAT.
There we argued that there were two classes of ways one might constrain the
parameters in a theory. Let us illustrate these two classes with the aid of
Figure 2. First, we could require $\Q$ to be such that the probability density
of the observed value of $\Delta$ is reasonably high. For example, if
$\Delta$ were $0 ~\mu$K, we might allow the parameters $\Q=10\mu$K and 
$\Q=20\mu$K, but frown on $\Q=30\mu$K, because the probability density of
$\Delta=0~\mu$K is unacceptably low. If, however, $\Delta$ was observed to be
$200\mu$K we would rule out all three values of $\Q$, because the probability
density is too low in each case. Now consider the second method: the likelihood
approach. In this approach, we compare the probability density of $\Delta$ for
different values of $\Q$ and throw out values of $\Q$ which have likelihoods
significantly smaller than the ``best'' values of $\Q$. In our artificial
example here, with only three values of $\Q$, this means that if $\Delta$ were
$200\mu$K, we would throw out $\Q=10,20\mu$K, but keep $\Q=30\mu$K since
this is the value of $\Q$ at which the likelihood [for the observed $\Delta$]
is maximum. There is something unsatisfactory about this: We are accepting a
value of the parameter which gives the best fit, but it is not a particularly
good fit. So although we will follow the second approach to find the best fit
$\Q$, we will use the first approach to ``check'' the goodness of this best
fit.
\putfig\pthree{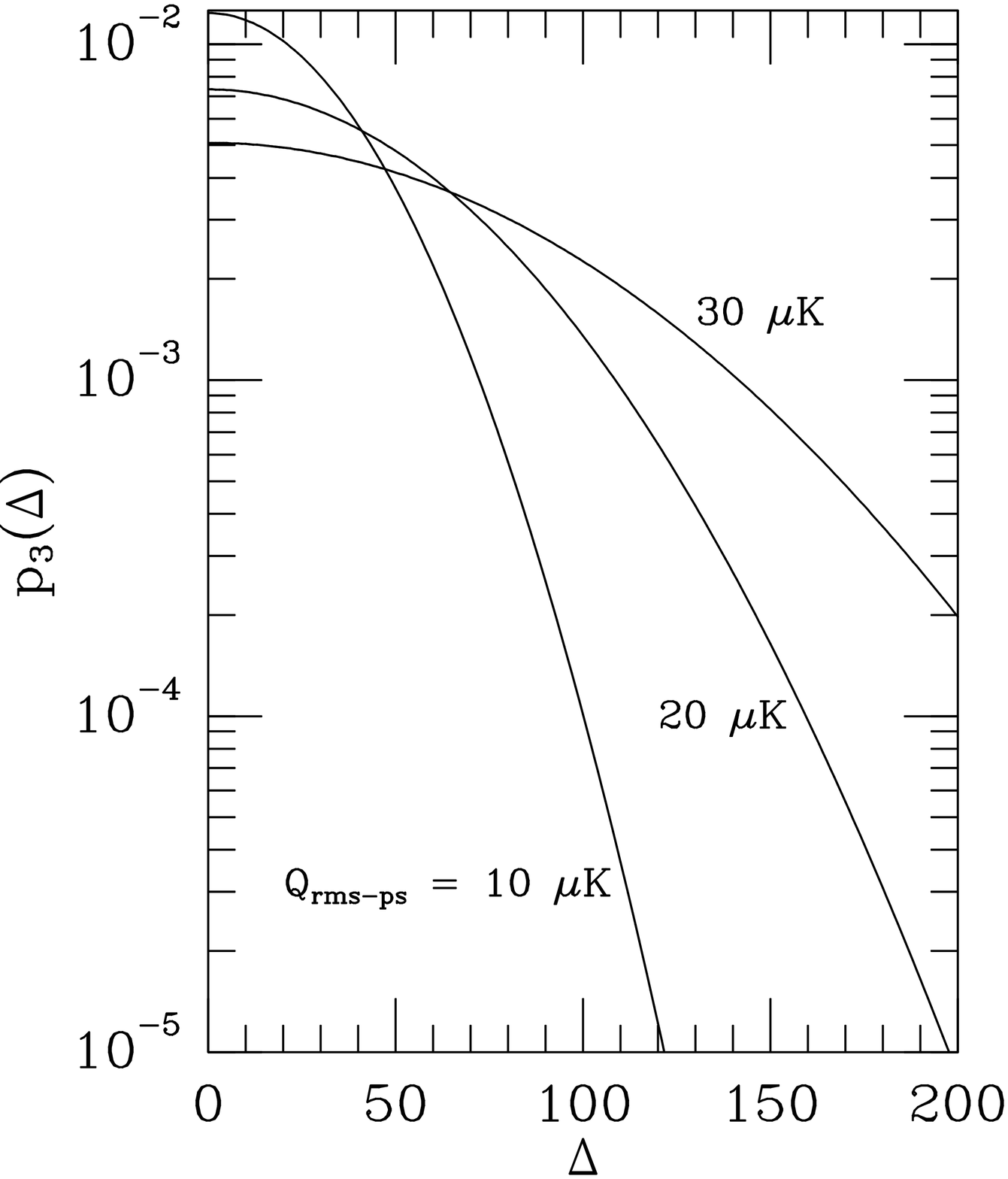}{The probability distribution for the observed
temperature in a single patch of the SP91 experiment, $p_3(\Delta\vert \Q)$,
for three different values of the CDM normalization, $\Q$. The distribution,
which is the same for $-\Delta$, is normalized so its integral over all
$\Delta$ is one.}

\subheader{\bf Many Patches; One Frequency Channel}

The nine point scan of SP91 reported temperature differences for nine patches
on the sky. The discussion above is easily generalized to the multi-patch case.
The data $\cal D$ which before was a single measurement $\Delta$ is now a
series of measurements, $\Delta_i, i = 1, \ldots N_{\rm patch} = 9$. We have
seen that all we need to calculate $p_3$ [or the likelihood function] is the
correlation matrix, $C$, which now is $9\times 9$. The experimental errors are
assumed independent so
\ben
C_{ij} = <T_i T_j> + \delta_{i,j} \sigma^2_{{\rm exp},i} .\ee
The off-diagonal elements of the theoretical correlation matrix $<T_iT_j>$ are
given by the same sum in eq.~\ELSUM, with
\be\SPFIL W_{l,ij} = \exp\left\{- l(l+1)\theta_s^2 \right\}
	{16\pi\over 2l+1} \sum_{m=-l}^l \cos(m(\phi_i-\phi_j)) H_0^2(m\phi_A)
Y_{lm}^2(\theta_z,0)
\ee
where $\phi_i$ is the azimuthal angle corresponding to the center of the
$i^{th}$ patch [the polar angle $\theta_z$ is kept constant throughout the
scan].

There is one further complication that must be dealt with before we can show
the likelihood function. The experimenters subtract from each scan an
``offset'' and a ``drift.'' That is, they subtract from each scan the best fit
line. So the temperature differences they are actually reporting are
\ben
\Delta'_i \equiv \Delta_i - (m \phi_i + b).
\ee
There are several ways of accounting for this. Bond {\it etal.} (1991) assumed
a uniform prior for the average and gradient and integrated them out. Another
possibility is to note that the $8^{th}$ and $9^{th}$ patch are really
redundant, since they are fixed by the requirement that the mean and gradient
vanish. Thus we could simply project onto the seven-dimensional space of
measurements. All of this has a familiar ring to it: these are precisely the
two alternatives we found to be equivalent when we talked about foreground. So
as a warmup exercise to subtracting off foreground, let's apply the formalism
we set up in section $2$ to subtract off the mean [we won't worry about the
gradient in this discussion, although we do subtract it off to get our final
results]. Analogous to eq.~\NULLSPACE\ we want the final temperature set to
have zero mean. If all the errors were the same, this would mean
$\sum_{i=1}^{\npat}z_i=0$. Putting in the weighting factors leads to
\ben
\sum_{i=1}^{\npat} {z_i\over \sigma_{{\rm exp},i}^2} = 0.
\ee
This equation is satisfied for eight linearly independent $z_i^{(r)}, r =
1,\ldots , 8$. Once we find these eight vectors, we can then form the
mean-subtracted temperatures
\ben \Delta'_r = \sum_{i=1}^9 z_i^{(r)} \Delta_i . \ee
To calculate $p_3$, we can still use eq.~\EPTHREE, but now the correlation
matrix that enters is the reduced correlation matrix
\ben C_{rs} \equiv \langle \Delta'_r\Delta'_s \rangle \
		= \sum_{i,j=1}^{\npat} z_i^{(r)} z_j^{(s)} C_{ij}.
\ee

\putfig\pfive{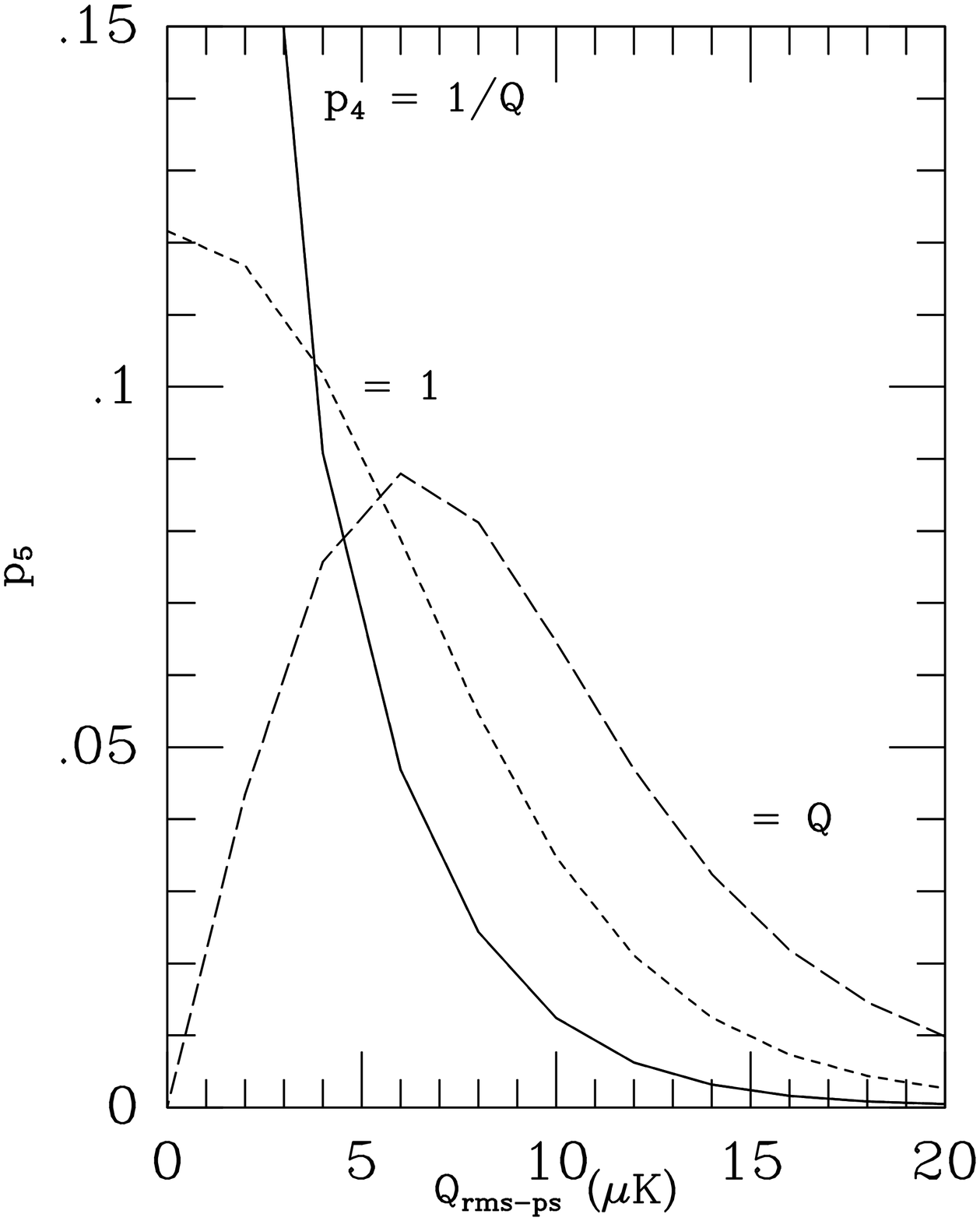}{The posterior distribution for $\Q$ using the highest
frequency channel of the SP91 experiment. The three different curves correspond
to three different choices of priors: $p_4 = 1/\Q,1,\Q$. The first
such prior has been cut off at $\Q=2\mu$K to make it normalizable.}

	If we assume a prior distribution function, $p_4(\Q)$, we can convert
$p_3$ into a posterior distribution for $\Q$, $p_5(\Q\vert\calD)$, using
eq.~\BAYES. (We see from eq.~\BAYES\ that the functional
dependence of the likelihood function on the model parameter is the same as the
posterior probability density for the parameters in a Bayesian analysis if the
assumed prior is uniform in the parameters. Since others (e.g.  Srednicki \etal
1993) have put a Bayesian interpretation on their analysis we will find it
convenient to discuss various prior distributions, and use the resultant
posterior probability densities to set limits on $\Q$.  However one should
remember that when we consider a uniform prior, we might just as well be
doing a likelihood analysis as a Bayesian analysis.  The names are changed but
they are mathematically equivalent.)
Figure 3 shows this posterior distribution function for the highest
frequency (4th) channel of the SP91 experiment, in the case of three different
priors. For the 4th channel SP91 data, no matter which prior is used , $p_5$
falls off fairly quickly, which led a number of groups to place fairly
stringent upper limits.  The exact limit does depend sensitively on the prior
though. If we define an upper limit via 
\be\uptest
\int_0^{\qup} dQ\,p_5(Q)=0.95 , 
\ee
then the upper limits are $\qup=9,\,14,\,20\mu$K for $p_4(\Q)=1/\Q,\,1,\,\Q$,
respectively.  One way to assess the upper limit associated with a given prior
is to calculate the {\it level of significance} of the test. This tells us how
often we would rule out $\qup$ if that was the true value. For example,
when $p_4(\Q)=1$ leading to an upper limit of $\qup=14\mu$K, we can ask: If the
Universe truly had $\Q=14\mu$K, and many different experiments were done, how
often would someone using this test rule out $\Q=14\mu$K? The levels of
significance of the tests with the priors $p_4 = 1/\Q,1,\Q$ are
$.11,\,.02,\,.002$, respectively. Thus a prior which favors high values of $\Q$
leads to very low levels of significance; i.e. it leads to upper limits
which are too stringent. For this reason, $p_4=1$ is often
chosen, and we will stick to this choice for the rest of our discussion.

\subheader{\bf Many Patches; Many Frequency Channels}

We now account for the fact that SP91, like many modern anisotropy experiments,
took measurements at a number of different frequency channels. In the SP91
experiments there were four frequency channels, spanning the range $25-35$ GHz.
Thus, the data set $\cal D$ now consists of $\Delta_{(a,i)}, a = 1, \ldots, 4;
i= 1,\ldots,7$ [recall that the mean and gradient are subtracted out of each
channel]. To construct the posterior distribution $p_5$ [from now on we'll use
$p_4 = 1$], we must therefore form the correlation matrix
\ben
C_{(a,i)(b,j)}=\langle T_{a,i}T_{b,j}\rangle+\delta_{ij}\delta_{ab}
                                                 \sigma_{{\rm exp},(a,i)}^2\ .
\ee
Ordinarily, the expected cosmic signal would be the same in each channel, since
the temperature differences are frequency independent. In SP91, there is a
small complication owing to the different widths of the beams in the different
channels.  We have accounted for this by allowing $\theta_s$ in the filter
function of eq.~\EFIL\ to be channel dependent.

Figure 4 shows the posterior distribution for $\Q$ given the 4-channel data in
the SP91 experiment. Taken at face value, this data clearly seems to indicate a
detection, i.e. $\Q=0$ is ruled out. Applying the test in eq.~\uptest\ to
determine an upper limit and a similar one to determine the lower limit, i.e.
\ben
\int_{\qlow}^{\infty} dQ\,p_5(Q) = 0.05 ,
\ee
we find
\be
\gfourbf \Q = 10^{+11}_{-4} \mu\rmK
\ee
at the $95\%$ confidence level.  Eq.~\gfourbf\ tells us that the best fit for
this experiment is at about $\Q=10\mu$K. Is this ``best fit'' a good fit? When
there was only one data point we could simply look at the distribution $p_3$
and see whether or not the distribution function was acceptably high at the
actual values observed.  Unfortunately, it is harder to do this in a 28
dimensional space. Instead, a good number to look at for these purposes is the
$\chi^2$, defined in eq.~\CHIIND.  The observed $\chi^2$ should be of order the
number of degrees of freedom, in SP91 $(4\ {\rm channels})\times(7\ {\rm
patches})-(1\ {\rm normalization\ parameter})=27$, with a standard deviation of
$\sqrt{2\times N_{\rm dof}}=7$.  However, in this case the observed $\chi^2 =
46$ for the four-channel data. This tells us that our best fit is not a very
good fit at all, and we better look at a different theory or at least at the
possibility of other sources. We turn next to this latter possibility.

\putfig\gaiernof{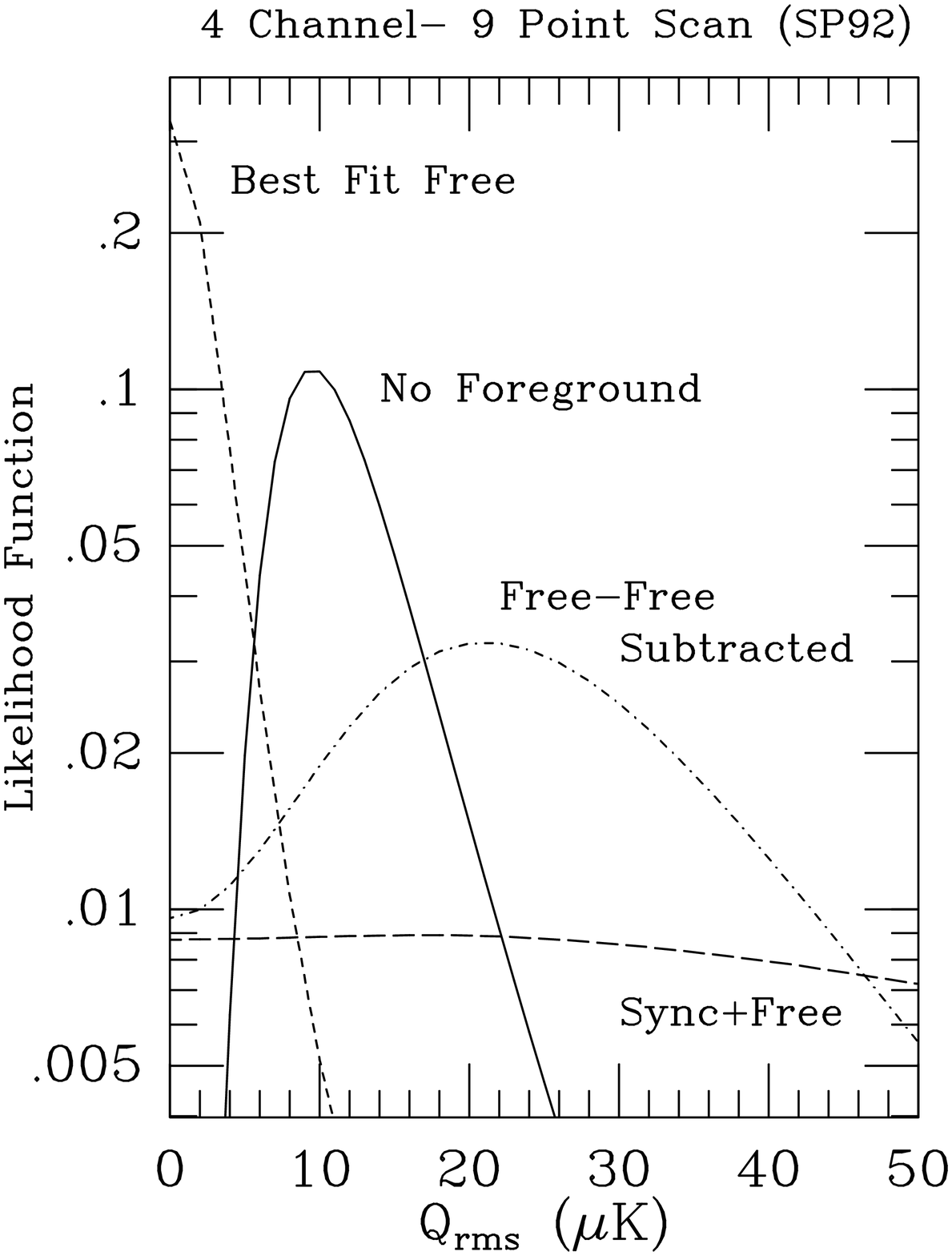}{The posterior distribution for $\Q$ from the
four channel, nine patch scan of SP91. The solid line assumes no foreground
contribution. The dashed line assumes a best fit free-free emission with
frequency dependence as in eq.~(34). Dotted line shows the result if free-free
emission is subtracted off. Dot-dashed line shows the results if synchrotron
and free-free are subtracted off.}

\subheader{One Source of Foreground: Marginalization}

We now allow the possibility of foreground sources. The observed signal then is
the sum of detector noise, foreground sources, and cosmic signal. As in
Eq.~\LIKE, the distribution function for the data is
\ben
p_3(\calD\vert\Q)={1\over(2\pi)^{(N/2)}}\vert\vert C \vert\vert^{-1/2}
\int d\calF\ p_6(\calF)\exp\left\{-{1\over2}(\calD-\calF)C^{-1}(\calD-\calF)
                                  \right\} .
\ee
The number of independent measurements $N=28$, one for each patch and channel.
Until now we have been implicitly assuming a delta function for the foreground
prior: $p_6 \propto \delta({\cal F})$, no foreground. In principle, $\calF$ is
a set of $28$ numbers, one for each patch and channel. Let us first suppose
that there is only one component to the foreground, free-free emission, say.
Then, we expect the signal in each channel to scale as 
\be\forenud
F_a \propto \nu_a^{-2.1}
\ee
where $\nu_a$ is the frequency associated with the $a^{th}$ channel. The
distribution function now becomes
\bea\foreint{
p_3({\cal D} \vert \Q) &= {1\over (2\pi)^{(\nch\npat/2)} }  \vert\vert C
\vert\vert^{-1/2}\int \Pi_{i=1}^{\npat}(dA_i)\cr
&\qquad\qquad\times p_6(\{A_i\})
          \exp\left\{-{1\over2}(\Delta_{(a,i)}-A_i F_a)C^{-1}_{(a,i)(b,j)}
                               (\Delta_{(b,j)}-A_j F_b)             \right\} .}
If the foreground contribution is known in one of the channels, eq.~\forenud\
allows us to determine the contribution in all channels. Thus, instead of $28$
free parameters, we have added only seven. Other maps of the region, radio maps
or infrared maps, might give a reasonable prior $p_6$, so that the integral in
eq.~\foreint\ could be done and translated into a distribution for $\Q$. In the
absence of these, we will use a uniform prior for the $A_i$, which we've argued
is equivalent to considering only those linear combinations of the temperatures
which are independent of the free-free emission. Before we do this, though, it
is interesting to consider another possible approach to the integral in
eq.~\foreint. Besides the prior, the only dependence on the free-free emission
comes in the quadratic polynomial in the exponential. One could easily minimize
the argument of the exponential, thereby maximizing the distribution. If the
integrand is sharply peaked, then this ``best fit foreground'' should be a good
approximation to the distribution. Figure \gaiernof\ shows the likelihood
function obtained in this way. The best fit free-free leaves little room for
anything else! The lesson to be learned from this example is {\it not} that a
very stringent upper limit has been set. Rather, it is important to note the
difference that foreground sources can make. The likelihood for $\Q$ with best
fit foreground differs significantly from the likelihood with no foreground.
Since the foreground can make a world of difference and since we really don't
have much prior information about the foreground, one reasonable approach is to
keep only the parts of the signal that are independent of foreground, i.e. to
marginalize.

We subtract off components dependent on foreground by solving eq.~\NULLSPACE;
we'll do this first assuming a foreground spectrum as in eq.~\forenud. The four
frequency channels in the SP91 experiment are centered at 
$\nu=26.25,28.75,31.25,33.75\,$GHz. Therefore, we must solve
\ben
\sum_{a=1}^4 z_a \left( {26.25\over 26.25 + (a-1)2.5}\right)^{2.1} = 0\ .
\ee
There are three independent solutions to this equation. So the four
temperatures in the four channels have been reduced to three temperatures
--linear combinations of the four channels:
\bean{
\Delta^{\rind}_{\s(1,i)}=&-2.5\Delta_{\s(1,i)}+ 1.3\Delta_{\s(2,i)}
                         +1.2\Delta_{\s(3,i)}+ 1.0\Delta_{\s(4,i)}         \cr
\Delta^{\rind}_{\s(2,i)}=& 0.0\Delta_{\s(1,i)}- 6.7\Delta_{\s(2,i)}
                         +9.6\Delta_{\s(3,i)}- 1.9\Delta_{\s(4,i)}         \cr
\Delta^{\rind}_{\s(3,i)}=& 0.0\Delta_{\s(1,i)}- 2.2\Delta_{\s(2,i)}
                         -0.7\Delta_{\s(3,i)}+ 3.9\Delta_{\s(4,i)}.
}
The index $i$ here still labels patches, as we perform this transformation in
each patch. The normalization is arbitrary; we have chosen it so that for MBR,
with its flat spectrum, $\Delta^{\rind} = \Delta$. From here on in, the
construction of the likelihood function mimics the process we went through with
mean-subtraction. The main step is to construct the reduced correlation matrix
as in eq.~\CORRIND. Figure \gaiernof\ shows the likelihood function thus
obtained.  Perhaps the most surprising feature of the marginalized likelihood
function is that it leads to a very weak upper limit:
\be\WEAK
\qup=52\,\mu\rmK\ .
\ee
This is a bit surprising because one might have reasoned as follows: Perhaps
the lower limit coming from the four channel data with no foreground should be
ignored because part of the signal could have come from something else. But at
least the upper limit should be reliable even if foreground is accounted for.
We see now that this reasoning is wrong! Not only can foreground confuse us
into thinking there is a cosmic signal when in reality none exists, but also
foreground can take away part of a cosmic signal. Foreground can be negative
when the cosmic signal is positive, so by ignoring foreground, one can 
{\it underestimate} the cosmic signal, and it is this possibility that leads to
the weak upper limit.  Some may find the weak upper limit of eq.~\WEAK\ overly
conservative since they would consider the possibility of the foreground
emission decreasing the variance in channel 4 to be 
negligible.   We will argue
against this point of view in below in \S5.

\subheader{Several Sources of Foreground}

Until now we have allowed only one source of foreground. What happens if we
allow several sources? The answer is disquieting. We now show that with 
more than
one source of foreground and no prior information about the amplitudes, no
interesting limits can be placed on the parameter $\Q$.

If we subtract a synchrotron component [assuming $F\propto \nu^{-2.7}$] in
addition to the free-free component, the two independent temperatures in each
patch are
\bea\twog{
\Delta^{\rind}_{\s(1,i)}=&11.3\Delta_{\s(1,i)}-21.8\Delta_{\s(2,i)}
                         -0.8\Delta_{\s(3,i)}+12.3\Delta_{\s(4,i)}         \cr
\Delta^{\rind}_{\s(2,i)}=& 0.0\Delta_{\s(1,i)}+21.5\Delta_{\s(2,i)}
                        -54.7\Delta_{\s(3,i)}+34.2\Delta_{\s(4,i)}  .
}
If three components -- free-free,synchrotron, and dust [$F\propto \nu^{1.6}$]
--
then the one independent component in each patch is
\be\threeg
\Delta^{\rind}_{\s(1,i)}=
116\Delta_{\s(1,i)}-420\Delta_{\s(2,i)}+494\Delta_{\s(3,i)}-189\Delta_{\s(4,i)}
.\ee
Recall that $\Delta^{\rind}$ is the sum of the cosmic component $(T^{\rind})$
and the detector noise $(N^{\rind})$. As we've said, the cosmic component
contributes equally in each channel so in a given patch $T^{\rind} = T$.
However, the detector noise is completely uncorrelated in the different
channels, so the large coefficients in Eqs.~\twog\ and \threeg\ tells us that
the detector contribution to $\Delta^{\rind}$ will be enormous. The cosmic
signal will be dwarfed by the detector noise. As an example of this, Fig.
\gaiernof\ shows the likelihood function when both
synchrotron and free-free are
subtracted off. The likelihood function is extremely flat, and the upper limit
is not very useful: $\qup = 200\mu$K.

These conclusions are not particular to SP91. For all the experiments we will
analyze, subtracting off more than one source of contamination results in data
swamped by noise and therefore useless for setting limits. This may be due to
the small number of channels [three for MAX-ACME and four for the South Pole
scans]; perhaps more channels would improve the situation. And of course we
have not used any information from other maps which might further constrain the
sources.

Perhaps subtracting off one source is enough though. One of the main incentives
for considering foreground when analyzing the SP91 results was the large
$\chi^2$ for the best fit $\Q$ using the four channel data [$\chi^2 = 46$ for
$27$ dof]. When only one source of foreground is included $\chi^2= 28$ for $20$
dof, a significantly better fit. In all of the experiments we analyze here,
subtracting off one foreground component makes the best fit $\Q$ a better fit.

\global\advance\countsec by 1

\header{\number\countsec. Other Experiments}

In this section we apply the method developed and explained in the previous two
sections to three other recent experiments. Along with the nine point scan of
the South Pole experiment, there was a $13-$ point scan (Schuster, \etal 1992)
which probed a nearby region of the sky.  At larger frequencies, the Millimeter
Anisotropy Experiment (MAX) recently reported results of two scans: one around
the region of $\mu-$Pegasus and the other near the star Gamma Ursa Minoris
(GUM).

\putfig\rest{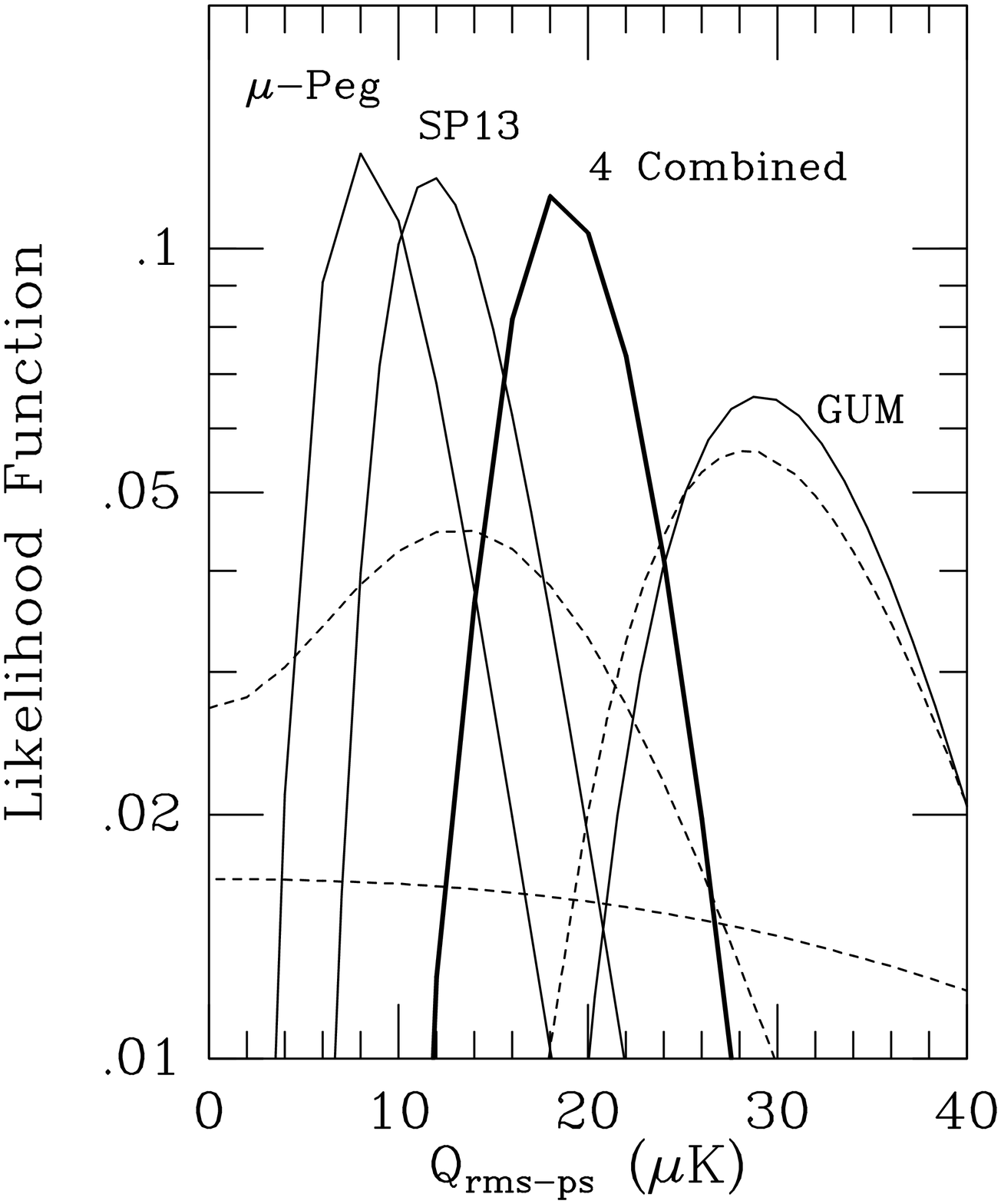}{The likelihood functions for the SP 13 point scan
and two MAX scans. The solid line labeled $\mu-$Peg is the likelihood for
$\mu-$Peg when warm dust is marginalized; the bottom-most dashed line is for
$\mu-$Peg when both warm dust and free-free [with $T \propto \nu^{-2.1}$] are
marginalized. The solid line labeled SP13 is for the raw data presented in the
thirteen point scan of the South Pole experiment; the dashed line which is
broader but peaks in roughly the same place is for the same experiment with
free-free marginalized. The solid curve for the GUM scan is the likelihood
using the raw data; the nearby dashed curve shows the likelihood when cold dust
is marginalized. The heavy solid line is the product of the likelihood
functions for the four experiments, each with one foreground component
marginalized.}

%\nopagenumbers
%\magnification1200
\begintable
Experiment | Foreground \|   $Q_{\rm rms}/\mu$K | $\chi^2$/d.o.f. |
$P(\chi^2)$ \crthick
SP 9-point scan | None  \|    $10^{+11}_{-4}$ | $46/27$ | $.012$ \cr
		| Free-Free \| $22^{+25}_{-17}$ | $28/20$ | $.12$ \cr
		| Free-Free + Syncrotron \| $<184$ | $18/13$ | $.16$
				\cr
SP 13-point scan | None \| $12^{+8}_{-4}$ | $58/43$ | $.06$ \cr
		| Free-Free \| $14^{+16}_{-12}$ | $40/32$ | $.17$\cr
MAX: $\mu-$Peg scan 	| Warm Dust  \| $8^{+9}_{-4}$ | $35/41$ | $.73$ \cr
			| Warm Dust + Free-Free \| $<86$ | $19/20$ | $.52$ \cr
MAX: GUM scan | None \| $29^{+14}_{-7}$ | $149/113$  | $.02$ \cr
		| Cold Dust  \| $29^{+16}_{-9}$ | $98/75$ | $.05$ \crthick
All Four Experiments | One Component \| $18^{+8}_{-5}$ | $208/168$ | $.02$
\endtable
{\figfont \noindent Table 1. Results of Bayesian analyses of four experiments.
$\Q$ lists the best fit $\Q$ with the error bars indicating the $95\%$ upper
and lower limits using Bayesian analysis with a uniform prior for $\Q$. The
column headed $\chi^2/$dof lists the $\chi^2$ of the best fit $\Q$. The
last column tabulates the probability of getting a $\chi^2$ as large as
or larger than this for the stated number of degrees of freedom.}

Figure 5 shows the results of calculating the likelihood function for these
experiments with and without marginalization. Table 1 gives quantitative upper
and lower limits and presents the $\chi^2$ for the best fit value of $\Q$. Also
tabulated is the probability of getting a $\chi^2/$dof this large, which we'll
take as a measure of goodness-of-fit.

The raw Schuster data indicates a detection, but as Table 1 shows, the best fit
for this data is quite poor if we assume it is all cosmic background.
Specifically, the probability of getting a $\chi^2$/dof$=58/43$ or larger is
only $.06$.  The situation improves somewhat if one foreground component is
subtracted off. In this case, though, the detection becomes less significant:
the likelihood function at $\Q=0$ is still $60\%$ of its maximum at
$\Q=12\mu$K.

The MAX experiment has a slightly different chopping procedure than the South
Pole experiments, so Eq. \SPFIL\ is replaced by
\bea\MAXFIL{
W_{l,ij} &= \left({3.34\over 2}\right)^2 \exp\left\{-l(l+1)\theta_s^2\right\}
	{16\pi\over 2l+1} \sum_{m=-l}^l
	\left( {\sin(m\phi_{*,i}/2) \over m\phi_{*,i}/2} \right)
	\left( {\sin(m\phi_{*,j}/2) \over m\phi_{*,j}/2} \right)
\cr\vs
&\qquad\qquad\times\cos(m(\phi_i-\phi_j)) J_1(m\phi_{A,i}) J_1(m\phi_{A,j})
Y_{lm}^*(\theta_{z,i},0) Y_{lm}(\theta_{z,j},0).
}
Here $J_1$ is the Bessel function of order one. For both scans the beam
smearing angle $\theta_s = 0.425 \times 0.5^\circ$, and the chop amplitude was
$\phi_A \sin(\theta_z) = 0.65^\circ$. The $\mu-$Peg scan was taken at constant
$\theta_z = 65.45^\circ$, with an angle of $\phi_* =0.285^\circ/\sin(\theta_z)$
between the center of each spatial patch. The GUM scan was binned so that
$\phi_* = 1.125^\circ$. The scan around GUM took data at several strips; that
is, $\theta_z$ was not constant.  We account for this by allowing each patch to
have $\theta_{z,i}$, a polar angle which varies from row to row [there are four
rows in all]. Eq. \MAXFIL\ is based on the assumption that the scan took place
at constant $\theta_z$, whereas the scans really took place in a ``bowtie''
pattern. However, if the secondary chop was rapid enough, the constant azimuth
approximation should be a good one. And indeed our results for the raw data
alone agree with other preliminary analyses (Bond 1993; Srednicki, {it et al.}
1993), the latter of which did account for the bowtie pattern. The prefactor in
Eq. \MAXFIL\ is explained in Srednicki {\it et al.} (1993); it normalizes the
signal so that the MAX filter really would report $T_1-T_2$ if it scanned
across a region with two different temperatures.

The $\mu-$Peg data correlates very well with dust from the IRAS catalogue. The
MAX team used two methods to extract information about the MBR from this data.
First, the IRAS dust was directly subtracted, and residuals analyzed as MBR;
second, the full data was simultaneously fit for dust and MBR. Subsequent
analyses have used the residual data set. Both of these methods give similar
answers. It is worth pointing out though that marginalizing with respect to
dust [as we do in Figure 5] is yet another method: instead of subtracting off a
best fit dust component [as the simultaneous fit does] marginalizing integrates
over all possible dust contributions, weighting each by the internal data. This
third method, represented by the solid line in Figure 5, gives results in
striking agreement with the other two. The dotted line in Figure 5 [the most
horizontal one] shows what happens if two components are subtracted off. Again,
little information can be gleaned; the upper limit is $86\mu$K.

The likelihood for the GUM scan is also shown in Figure 5.  The solid line is
the likelihood for the raw data, which leads to a high normalization [recall
that COBE's normalization is now $\Q=17\pm 3 \mu$K]. Table 1 shows that this
best fit is not a particularly good fit; however marginalizing with respect to
cold dust leads to little change in either the best fit or the goodness of fit.

Finally the heavy solid line in Figure 5 shows the combined likelihood function
of all four experiments we have analyzed. Here we have simply multiplied the
four likelihoods together, in each case choosing the
one-foreground-component-subtracted data. The normalization agrees eerily well
with that of COBE, but this is not necessarily the number to focus on. Instead,
we note that the likelihood function thus obtained is quite narrow; Table 1
shows that $\Q=18_{-5}^{+8}\mu K$, so the error bars are quite small.  This is
encouraging, because a valid criticism of this work would be that we have been
too liberal in our assumption about the amplitude of the foreground. The small
error bars show that we have not thrown out too much information. This best fit
is also a poor fit: the probability of getting a $\chi^2/$dof$=208/168$ or
larger is only $0.02$ reflecting the fact that all of the individual
experiments except for the $\mu-$Peg scan exhibit relatively poor fits with
large $\chi^2$'s.  The generally poor fit might be simply telling us that the
theory we have chosen is not the correct one.  A theory with more power on
small scales would help fit the GUM results better. Nonetheless, the fact that
the two MAX scans sample the anisotropy at the same angular scales and find
very different results suggests that it will not be easy to fit the data by
simply adjusting the angular power spectrum.  The probable interpretation is
that foreground has still not been adequately removed from the data.  Perhaps
the foreground has a very different spectrum than has been assumed here, say
self-absorbed synchrotron or very cold dust. Of course we may have assumed not
only the wrong power spectrum for the MBR but also the wrong statistical
distribution. If the primordial anisotropy field is not Gaussian then one might
expect a greater probability of having very different levels of anisotropies in
one part of the sky than in another.  This might explain the difference between
the $\mu-$Peg and the GUM results.  Here it should be remembered that our
analysis of the different experiments includes the uncertainty due to finite
sampling. One cannot ascribe the apparent inconsistency between GUM and
$\mu$-Peg under the CDM model to finite sampling.

\global\advance\countsec by 1

\header{\number\countsec. Are Our Limits Too Conservative?}

	In the preceding discussion we have taken into account the possibility
of foreground emission by throwing away those components of measurements which
might be effected by the sources of foreground we are considering.  This
process of marginalization makes no attempt whatsoever to determine the amount
of foreground contamination from the microwave data themselves.  As illustrated
by both fig.~\pfive\ and the table, this procedure can indeed lead to very weak
limits on the parameters characterizing the primordial anisotropy.  Is throwing
out all of this data really justified?

	Clearly the limits produced by marginalization are conservative ones,
as it allows for foreground contamination with an amplitude so large that it is
unlikely to have produced the data that is being analyzed.  One could
reasonably try to limit the foreground contamination by insisting on
goodness-of-fit to the microwave data and then using the constraints on the
foreground to construct a prior for the foreground emission, i.e. $p_6$ of
eq.~\LIKE.  The marginalization procedure is simpler and more straightforward
in that no spatial correlation for the foreground emission has to be
constructed.  In many cases marginalization will not lead to much less
stringent limits than those obtained by the Bayesian approach just described.
The utility of marginalization is best considered in light of the commonly used
alternative procedures:
\vskip-\parskip
\item{1)} analyze ignoring the possibility of foreground contamination, 
\vskip-\parskip
\item{2)} analyze ignoring foreground after removing the best fit foreground, 
or
\vskip-\parskip
\item{3)} analyze after culling of data points which appear to be contaminated.

\vskip-\parskip
\noindent All of these procedures we would consider to be dangerous as we will
now explain. 

	A good guiding principle in dealing with foreground is that one treat
the detector noise and ``foreground noise'' (i.e. contamination) on an equal
footing.  If we have no good information about the foreground contamination
from radio or IR measurements the only real differences between the two is that
one has a good idea of the characteristics of the former and very little idea
of the characteristics of the latter.  In the MBR community it has been
generally accepted that one should not use data whose amplitude is less than
the sensitivity (i.e. noise level) to put limits on the MBR anisotropy below
the sensitivity of the detector.  We would like to generalize this to say that
one shouldn't set upper limits on primordial anisotropy below the noise level,
and one must include in this noise the uncertain contribution from foreground
contamination.  If one has low detector noise and good spectral coverage one
can determine the foreground contribution accurately, and thus there is little
foreground uncertainty or noise.  However if one cannot determine the
foreground contribution accurately then the foreground noise is large and one
should account for this in setting limits.  This philosophy then leads us to
reject both alternatives 1) and 2), which take no account of the added
uncertainties due to foreground.  Alternative 2) is especially dangerous since
one not only underestimates the uncertainties but if the uncertainties in the
foreground are large, one may also subtract away much of the MBR signal.  This
is illustrated for the SP91 experiment in fig.~\pfive.  The small spectral
coverage of the experiment allows for very little discrimination between
free-free emission and primordial anisotropy, and subtracting the best fit
free-free signal therefore will lead to the subtraction of much of any
primordial anisotropy present.

	Of course, the problem with alternative 3) is determining which points
are likely to be contaminated in a way which is unrelated to the amplitude of
the primordial anisotropy at those points.  If ones criterion for culling a
data point has anything to do with the amplitude of the signal obtained at that
data point, then one runs the risk of biasing the limits on primordial
anisotropy.  One proper way of culling data from multi-frequency experiments is
to drop all of the data in patches where the signals which are linearly
independent of primordial anisotropies are particularly high.  Such linear
combinations are found by solving equations analogous to eq.~\NULLSPACE.

	In many analyses of the SP91 experiment, only the 4th channel is used,
essentially culling data from channels 1-3.  One would hope that the rationale
for this comes from the {\it a priori} knowledge that experiments in this
frequency range are most likely to be contaminated by free-free and synchrotron
emission and therefore channel 4 will be the least contaminated.  Unfortunately
the signal in channels 1-3 is significantly greater than that in the 4th
channel, which essentially meant that the data that had significant signal was
dropped and that which had none was kept.  Since the channels are so close in
frequency the signals in channels 1-3 give us a strong indication that whatever
is causing this signal, and it might be largely primordial anisotropy, should
also be present in channel 4 since only an implausibly steep spectrum would
give a negligible contribution to this highest frequency channel.  We would
therefore argue that the uncertainty in the foreground contamination of channel
4 is quite large and therefore any small upper limit on $\Q$ derived from
analyzing channel 4 while ignoring this foreground noise is unreliable.

	One might argue that upper limits such as that of eq.~\WEAK\ are too
weak since the limit is weaker than that obtained by considering only channel 4
and ignoring foreground contamination.  The rationale for this argument is that
it is much more probable that the foreground will increase the observed signal
rather than decrease it.  We do not find these arguments very convincing.
After all, given only 7 degrees of freedom the probability that $\chi^2$ is
less than half its expected value is $0.165$, not a very small number.  It is
interesting to note that if, say, half the signal comes from primordial
anisotropy and half the signal comes from foreground contamination, then a low
$\chi^2$ is much more likely to be produced by a cancellation of the foreground
and the primordial anisotropy than by having the foreground and primordial
contribution both be small.  This is easily understood in terms of phase space
arguments as follows. Consider two random variables $X$ and $Y$ both uniformly
distributed in $[-1,1]$ (see Figure 6).  
Thus the probability distribution in the $X$-$Y$
plane is uniform in a square.  The locus of possible outcomes which have a
given value of $X+Y$ are just lines that cut through the square at a
$-45^\circ$, and if this sum is much smaller than 1 this line passes close to
the origin, i.e. close to the diagonal $X=-Y$.  Most of the length of such a
diagonal line is not located near the origin where both $X$ and $Y$ are small
but is rather closer to the corners where $X,\,Y\sim1$ and there is strong
cancellation.  Thus if $X+Y$ is small it is more likely that $X\approx-Y$
rather than both $X$ and $Y$ being small.  The same arguments works when a
Gaussian distribution replaces a uniform distribution.  To reiterate:
cancellation is likely where the total signal is low even though it is unlikely
in general.  One may apply this to channel 4 of SP91 where we think there is
evidence that the signal is low.  If both foreground and primordial anisotropy
are contributing to channels 1-3 we shouldn't be surprised to find significant
cancellation between primordial and foreground contamination in channel 4.

\putfig\square{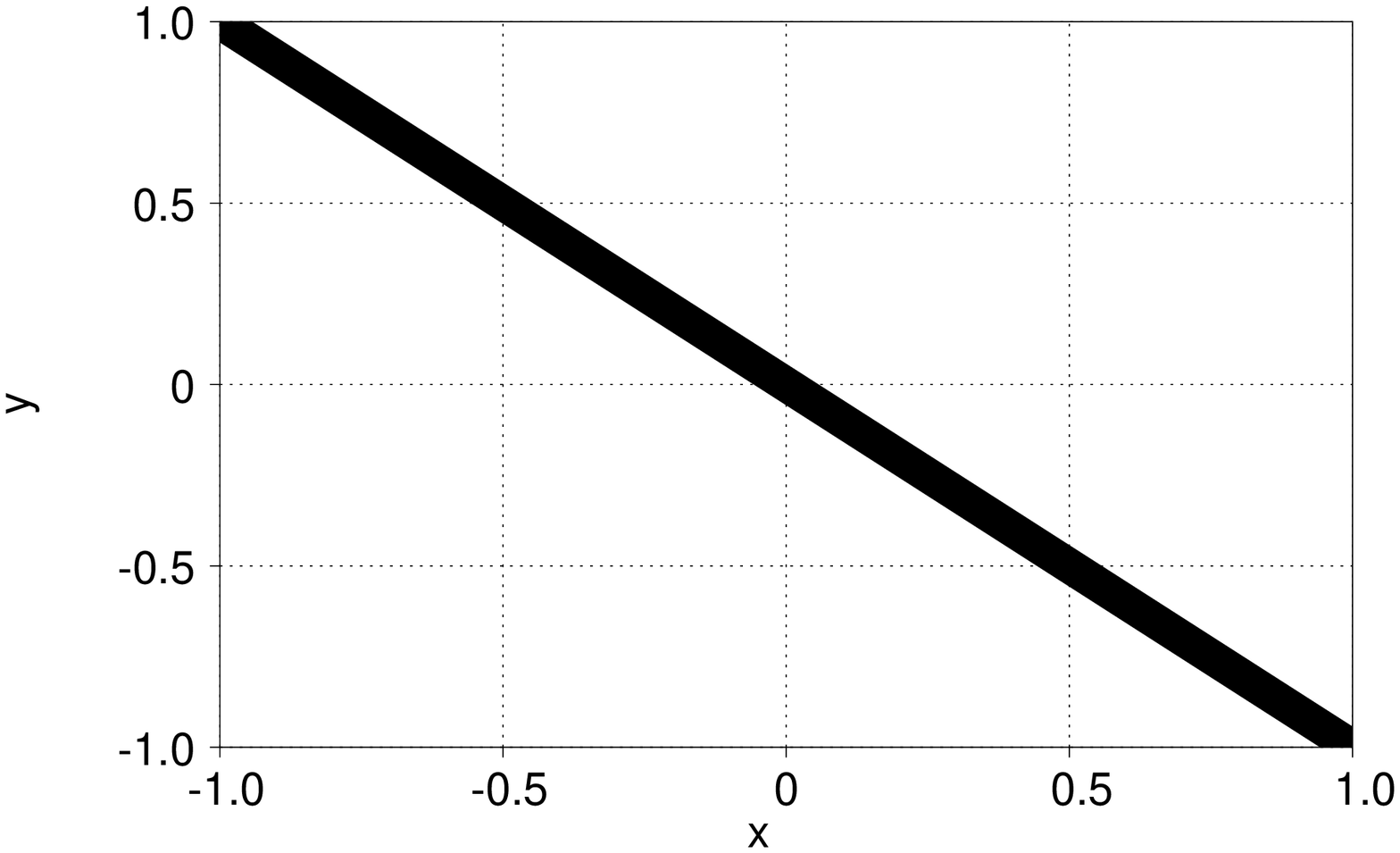}{A plane in which $X$ and $Y$ are randomly 
distributed. Even if data constrains $\vert X+Y\vert<.05$ [the shaded
region in the figure], neither $X$ nor $Y$ is necessarily small. In fact,
in much of the shaded region, both $\vert X\vert$ and $\vert Y\vert$ are of
order one.}

	In our analysis we have not made use of any other data outside of the
microwave anisotropy data to limit the amount of foreground contamination.  One
should be careful with such procedures as they often require bold
extrapolations of the spectrum. In contrast, to marginalize we require only
that the assumed form of the spectrum holds over a relatively narrow frequency
region [i.e. there is a big difference between extrapolating a $\nu^{-3}$
spectrum from $408$ MHz to $40$ GHz and extrapolating the same $\nu^{-3}$
spectrum between $25$ and $35$ GHz].  While there is often reliable physics
behind the extrapolations there are usually some caveats behind the
applicability of these extrapolations.  For example free-free or synchrotron
self-absorption may severely decrease emission in the radio of sources which
are bright in the microwave. Of course, if we have no idea about the spectra of
the foreground we cannot marginalize the signal from these sources.  Let us
hope that sources with such problematic spectra do not exist in any abundance.
In spite of the caveats we do take seriously these measurements. In many cases
the data indicates that there should be no significant contamination, and in
these cases marginalization may yield overly broad upper and lower limits.

	 In the future what must be hoped for is better data with low enough
noise in enough different frequency channels to be able to fit all of the
components to microwave brightness fluctuations.  We would like to 
point out that
the null space analysis of eq.~\NULLSPACE\ can be useful in choosing which
channels to use for such experiments.  Many of the very large upper limits in
the Table are simply a result of the coefficients obtained by solving
eq.~\NULLSPACE.  In other words, for some choices of channel frequencies the
signal-to-noise is greatly reduced by projecting out the foregrounds while for
other choices the reduction is less severe.  By choosing the right frequencies
one can optimize the signal-to-noise which is left after marginalization.  Of
course, instrumental considerations as well as considerations of atmospheric
emission must also play a strong role in the choices of frequencies.

This work was supported in part by the DOE and NASA grant NAGW-2381 at
Fermilab. 

\vfill\eject

\header{REFERENCES}
\def\refindent{\par\hangindent=.725 truein \hangafter=1\ }

\refindent Bond,~J.R., Efstathiou,~G., Lubin,~P.M., and Meinhold,~P.R. 1991,
PRL 66, 2179 

\refindent Brandt,~W.N., Lawrence,~C.R., Readhead,~A.C.S., Pakianathan,~J.N.,
and Fiola,~T.M. 1993, CalTech-OVRO preprint \# 12

\refindent Gaier,~T., Shuster,~J.,  Gundersen,~J., Koch,~T., Meinhold,~P., 
Sieffert,~M., and Lubin,~P., 1992, ApJ 398, L1

\refindent Gundersen,~J., \etal 1993, ApJ 
413, L1 

\refindent Meinhold,~P., \etal 1993, ApJ 409, L1 

\refindent Myers~S.T. 1990, CalTech Ph.D. thesis

\refindent Meinhold,~P. and Lubin,~P. 1991, ApJ 370, L11

\refindent Schuster,~J., Gaier,~T., Gundersen,~J., Meinhold,~P., Koch,~T.,
Sieffert,~M., Wuensche,~S.C., and Lubin,~P. 1993, ApJ 412, L47

\refindent Smoot,~G.F., \etal 1992, ApJ 396, L1

\refindent Srednicki,~M., White,~M., Scott,~D., and Bunn,~E.T. 1993,
CfPA-93-TH-27 preprint

\end